\documentclass{aa}

\usepackage{txfonts}
\usepackage{graphicx}
\usepackage[colorlinks=true, allcolors=blue]{hyperref}
\usepackage{xcolor}
\definecolor{green} {rgb} {0.0,0.6,0.0}
\usepackage{placeins}

\begin{document}

\title{X-ray activity of nearby G-, K-, and M-type stars and implications for planet habitability around M stars\thanks{Table \ref{tab:sample} is only available in electronic form at the CDS via anonymous ftp to cdsarc.u-strasbg.fr (130.79.128.5) or via http://cdsweb.u-strasbg.fr/cgi-bin/qcat?J/A+A/.}}

   \author{E. Zhu\inst{1,2}
          \and
          T. Preibisch\inst{1}}

   \institute{Universit\"ats-Sternwarte M\"unchen,  Ludwig-Maximilians-Universität, Scheinerstr. 1, 81679 M\"unchen, Germany\\
              \email{preibisch@usm.uni-muenchen.de}
    \and IRAP, Université de Toulouse, CNRS, UPS, 9 avenue colonel Roche, BP 44346, 31028 Toulouse cedex 4, France     
             }

   \date{Received September 15, 2024; accepted December 12, 2024}

\abstract
   {The intense X-ray and UV emission of some active M stars has raised questions about the habitability of planets around M-type stars.}
   {We aim to determine the unbiased distribution of X-ray luminosities in complete, volume-limited samples of nearby M dwarfs, and compare them to those of K and G dwarfs.}
   {We constructed volume-complete samples of 205 M stars with a spectral type $\leq$\,M6 within 10\;pc of the Sun, 129 K stars within 16\;pc, and 107 G stars within 20\;pc. We used X-ray data from \textit{Chandra}, \textit{XMM-Newton}, eROSITA, and ROSAT to obtain the X-ray luminosities of the stars.}
   {Our samples reach an X-ray detection completeness of 85\%, 86\%, and 80\% for M, K, and G stars, respectively. The fractional X-ray luminosities relative to the bolometric luminosities, $\log(L_\mathrm{X}/L_\mathrm{bol})$, of the M stars show a bimodal distribution, with one peak at around $-5$, mostly contributed by early M stars (M0--M4), and another peak around $-3.5$, contributed mainly by M4--M6 stars. The comparison of the different spectral classes shows that 63\% of all M stars in our sample (80\% of the M stars with a spectral type $<$\,M4) have $L_\mathrm{X}/L_\mathrm{bol}$ values that are within the central 80\% quantile of the distribution function for G stars. In addition, 55\% of all M stars in our sample (and 72\% of the M stars with a spectral type $<$\,M4) have $L_\mathrm{X}/L_\mathrm{bol}$ less than 10 times the solar value.}
  {The X-ray activity levels of the majority ($\ge 60\%$) of nearby M dwarfs no later than M6 are actually not higher than the typical (80\%\;quantile) levels for G-type stars. The X–ray irradiation of habitable-zone planets around these stars should therefore not present a specific problem for their habitability.}


   \titlerunning{X-ray Activity of Nearby G-, K-, \& M-type Stars}
   \maketitle

\section{Introduction}\label{sec:intro}

M stars comprise about 70\% of all stars in the galaxy. Their small mass and radius are beneficial for planet detection, and numerous exoplanets have been detected in the last few decades, including some Earth-sized planets in the habitable zones of their host star \citep[e.g.,][]{proxima_cen, TRAPPIST1, LHS1140}. Like all late-type stars, M stars generally show magnetic activity
and consequently emit EUV and X-ray radiation \citep{rotation_period_mass2, M_dwarf_magnetic_activity2, M_dwarf_magnetic_activity,2023AJ....165..195B}.

A  convenient tracer of the level of magnetic activity is the ratio of X-ray versus bolometric luminosity, $L_{\rm X} / L_{\rm bol}$. For our Sun, this ratio is
about $6 \times 10^{-7}$ in soft X-rays\footnote{The Sun's soft X-ray luminosity in the $[0.1\!-\!2.4]\;{\rm keV}$ ROSAT band determined by \citet{Sun_Lx} for the full activity range over a typical solar cycle corresponds to activity levels in the range
$\log\left(L_{\rm X}/L_{\rm bol}\right) \simeq [-6.8 \,\dots\, -5.7]$ .}  \citep[see][]{Sun_Lx,alpha_cen_Lbol}, whereas some highly active M dwarfs show ratios of up to $\sim 10^{-3}$ or even higher. Such extreme levels of activity might cause severe atmospheric escape of the planets surrounding such stars, and could thus strongly limit the habitability of these planets \citep[see, e.g.,][for general reviews about habitability of planets around M dwarfs]{review, Airapetian_review}. While there are several prominent examples of highly active M dwarfs \citep[e.g.,][]{AD_Leo, EV_Lac, YZ_Cmi, 2023AJ....165..195B}, it is very important to keep in mind that not all M dwarfs show such strongly elevated magnetic activity.
Prominent examples of highly active M dwarfs that were detected as ``flare stars'' or in X-ray surveys easily catch attention, but they can also easily bias samples of M dwarfs by causing a selection effect that favors objects with particularly high activity. The only way to get objective and unbiased information about the activity levels of M dwarfs is by constructing spatially complete samples -- so-called ``volume-limited''  samples containing all M dwarfs up to a specified distance from the Sun. 

Until recently, the completeness  of the census of M dwarfs in the solar neighborhood was restricted to distances of just a few parsecs, with correspondingly small sample sizes. The availability of \textit{Gaia} data has strongly improved the situation and allows one to now construct much larger complete, volume-limited samples of M-dwarfs.
This provided the motivation to  conduct a new comprehensive study of M dwarf X-ray activity, to investigate the true, unbiased distribution of activity levels in M dwarfs, and to compare it to volume-limited samples of K and G stars.

\cite{NEXXUS} have constructed a complete X-ray catalog of nearby GKM stars based on ROSAT all-sky survey and pointed observations, and studied the X-ray luminosity of nearby stars with different spectral types (SpTs). The \cite{NEXXUS} database has been used for several studies (e.g., \cite{Preibisch05, NEXXUS_study, 2018MNRAS.479.2351W, NEXXUS_study2}).
In the last 20 years, many new X-ray data have become available due to the \textit{Chandra}, \textit{XMM-Newton}, and, most recently, the ROentgen Survey with an Imaging Telescope Array (eROSITA) missions, providing new data for unbiased studies of the X-ray activity of late-type stars.

In Sect.~\ref{sec: M_sample} of this work, we describe how the optical nearby G-K-M star samples were constructed and how the stellar bolometric luminosities were determined. Section~\ref{sec:X-ray} describes how stellar X-ray luminosities (and upper limits in the case of non-detections) were determined from various X-ray observatories, and Sect.~\ref{sec:discussion} discusses  the completeness of the final X-ray catalogs of this work. In Sect.~\ref{sec:LxLbol-GKM}, we analyze and compare the $L_\mathrm{X}/L_\mathrm{bol}$ distributions of the different spectra classes, and in Sect.~\ref{sec:habitability} we finally discuss their implications for planet habitability around M dwarfs.

\section{Construction of the stellar samples}\label{sec: M_sample}

The stellar parameters we need for our analysis are
the SpT and the bolometric luminosity ($L_{\rm bol}$)
of each star.
Combining these data with the determined X-ray luminosities ($L_{\rm X}$) (see Sect.~\ref{sec:X-ray})
allows one to calculate the fractional X-ray luminosity, 
 $L_\mathrm{X} / L_\mathrm{bol}$, which is not only a good tracer of the
 activity level of the star, but also provides a quantitative measure
 of the X-ray irradiation of planets\footnote{As $L_\mathrm{X} / L_\mathrm{bol} = F_\mathrm{X} / F_\mathrm{bol}$, and since a planet is in the habitable zone if it receives a bolometric stellar flux that is similar to the solar flux received by Earth, the ratio characterizes the level of X-ray irradiation of the planet.} around this star.

The stellar samples constructed in this work are based on the Fifth Catalog of Nearby Stars \citep[CNS5;][]{CNS5} and the 10~pc sample \citep{10pc_sample}, which are both volume-complete stellar catalogs based on \textit{Gaia} EDR3;  the former extends to 25~pc from the Sun, while the latter is restricted to 10~pc, but provides more detailed information on the individual stars. 

For all of the stars in the 10~pc sample, we used the stellar SpTs listed in that catalog.
For stars beyond 10~pc, we either
searched for them in the SIMBAD database\footnote{\href{https://simbad.cds.unistra.fr/simbad/}{https://simbad.cds.unistra.fr/simbad/}} \citep{SIMBAD} and collected the most recent reliable SpT  measurement (see Appendix \ref{appendix:reference}), or determined their SpT from their \textit{Gaia} $G$ band absolute magnitude $M_G$, using the stellar parameter table\footnote{\href{https://www.pas.rochester.edu/~emamajek/EEM_dwarf_UBVIJHK_colors_Teff.txt}{https://www.pas.rochester.edu/$\sim$emamajek/EEM\_dwarf\_UBVIJHK\_ \newline colors\_Teff.txt}} from \citet{EEM_table} (hereafter the Mamajek table).

As the aim of this work is to construct volume-limited samples with a high level of  completeness, the size of our samples is limited by the X-ray detection completeness, which will be discussed in 
Sect.~\ref{sec:completenss}. Our aim is to construct samples of G-, K-, and M-stars with similar sizes, in order to allow for a meaningful comparison of the distribution functions. The M star sample was restricted to stars not later than type M6, since later-type stars are so faint that the currently identified population might not be complete at 10~pc even with \textit{Gaia}. In this way, we constructed samples of 112 G stars within 20~pc, 139 K stars within 16~pc, and 220 M stars within 10~pc. 
The 15 M stars, 10 K stars, and 5 G stars that are  secondary components of X-ray non-resolved multiple systems had to be excluded from the final analysis. This leaves us with final samples of 205 M stars, 129 K stars, and 107 G stars.

The stellar bolometric luminosities, $L_\mathrm{bol}$, of most stars were taken from the \textit{Gaia} DR3 astrophysical parameter catalog  \citep{GAIADR3, GSP_Phot}.
For 29 G stars, 48 K stars, and 107 M stars in our sample without documented \textit{Gaia} $L_\mathrm{bol}$, we calculated their values by using the \textit{Gaia} bolometric correction (BC) routine, which calculates the BC in $G$ band ($BC_G$), as is described in detail in Appendix \ref{app: BC}.

\section{X-ray counterparts and luminosities}\label{sec:X-ray}

Our strategy was to always use the best available X-ray data for each individual star. In terms of sensitivity and angular resolution, \textit{Chandra} is in most cases the first priority, followed by \textit{XMM-Newton}. For stars that have not been observed with one of these two observatories, we used data from eROSITA and ROSAT.

In cases in which the X-ray properties of a target star had already been studied, we used the published X-ray luminosities.
For target stars without published X-ray properties, we downloaded the relevant X-ray datasets from the archives, and determined the count rates and X-ray luminosities (or, in the case of non-detections, upper limits to these quantities) directly from the data.

Many nearby stars show quite large proper motions, which must be accounted for in order to enable a proper identification with X-ray sources.
For the source identification in the X-ray images and the
 crossmatching with published X-ray source catalogs, we therefore always propagated the \text{Gaia} DR3 stellar coordinates to the epoch of the corresponding X-ray observation by using their \textit{Gaia} DR3 proper motions.

\subsection{\textit{Chandra}}\label{sec:chandra}

For 9 G stars, 5 K stars, and 24 M stars observed with \textit{Chandra}, we took their X-ray luminosities from published results (see Appendix \ref{appendix:reference}). If multiple observations of the same star are listed in the paper, the one with the longest observation time is taken. $L_\mathrm{X}$ of stars that have separated measurements during their quiet periods and flare periods are averaged.

For 2 G stars, 6 K stars, and 11 M stars that were observed with \textit{Chandra} but that do not have any published $L_\mathrm{X}$, we analyzed the archival data and searched for X-ray sources at the optical positions of the stars derived by propagating the \textit{Gaia} coordinates by the proper motions of the stars to the epoch of their \textit{Chandra} observation.
In all but one case (the G2IV star GJ~19, with an angular separation of $8^{\arcsec}$), we found  X-ray sources within $\leq 7^{\arcsec}$ of the optical position of the stars. In all cases, we inspected 2MASS \textit{J}-band images and confirmed that there are no other bright sources nearby with smaller separation. 
We used the command \texttt{srcflux} from the software CIAO \citep{CIAO} to extract their source flux from the data. The plasma temperature was set to 0.5~keV for M stars and 0.3~keV for G/K stars, and extinction was neglected. The observations used are listed in Table \ref{tab:chandra_observation}.

\subsection{\textit{XMM-Newton}}

For 7 G stars, 9 K stars, and 12 M stars observed with \textit{XMM-Newton}, we took their X-ray luminosities from published results (see Appendix \ref{appendix:reference}). Stars with multiple observations were dealt with using the same procedure that was described for \textit{Chandra}.

We then cross-matched our samples with the 4XMM\_DR13 catalog \citep{4XMM_DR13} with an initial search radius of $15^{\arcsec}$. We found matches for 34 M stars, 13 K stars, and 19 G stars; all these matches have positional separations of  $<7^{\arcsec}$, all but two $<5^{\arcsec}$.

For stars without pointed \textit{XMM-Newton} observations, we also employed the \textit{XMM-Newton} slew survey catalog \citep[XMMSL2;][]{xmmsl2}. \cite{XMMSL2_stellar} constructed a catalog of stellar counterparts of XMMSL2 by crossmatching it with \textit{Gaia} DR1, 2MASS, and \textit{Tycho2}, from which we collected the $L_\mathrm{X}$.  In the end, 2 M stars, 1 K star, and 2 G stars from our samples were assigned with $L_\mathrm{X}$ measured by XMMSL2.

\subsection{ROSAT}\label{sec:ROSAT}

For stars for which neither \text{\textit{Chandra}} nor \textit{XMM-Newton} data were available, we considered ROSAT data.

\subsubsection{NEXXUS}

\begin{figure*}[htbp]
     \centering
    \includegraphics[width=0.4\linewidth]{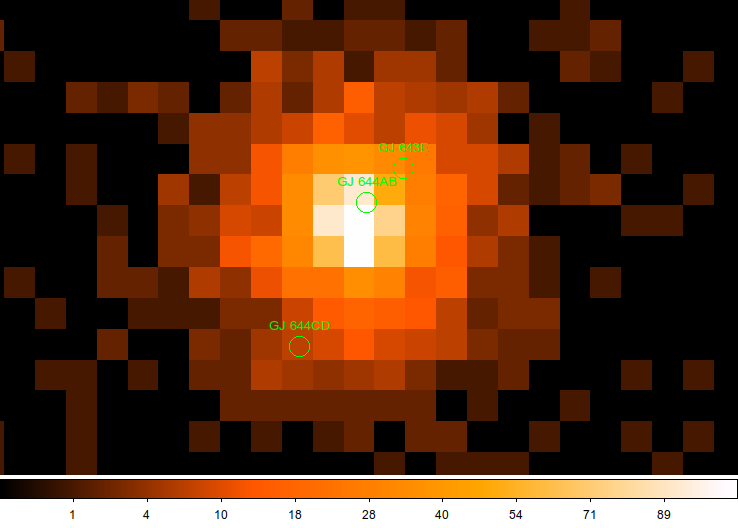}
    \includegraphics[width=0.4\linewidth]{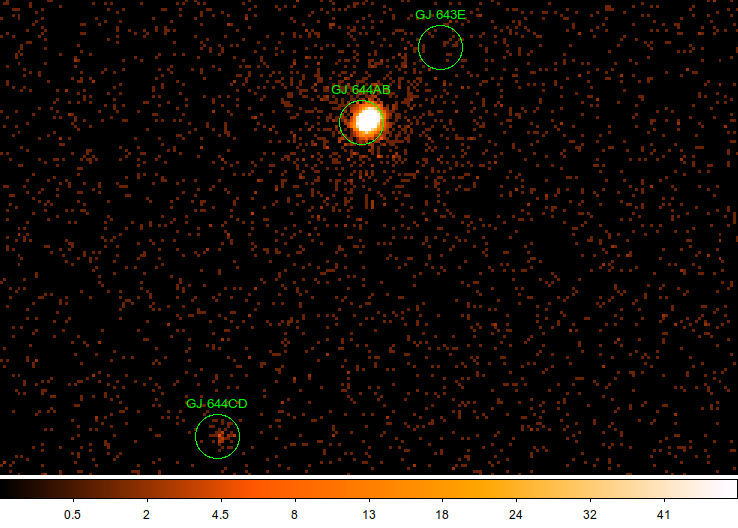}
    \includegraphics[width=0.4\linewidth]{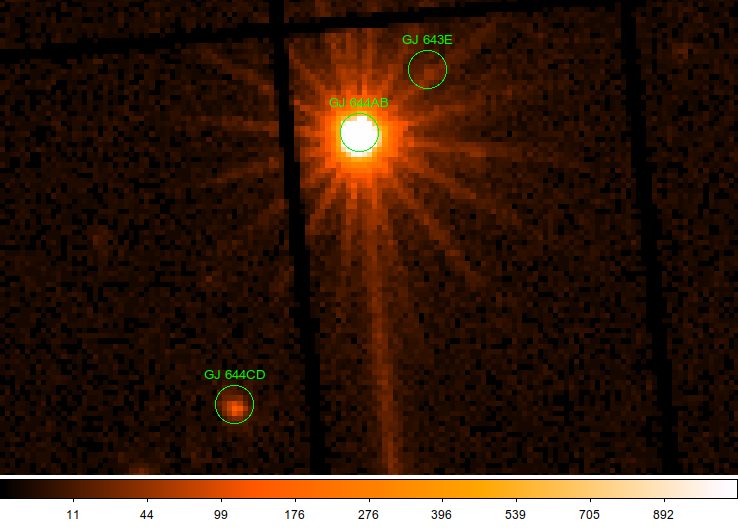}
    \includegraphics[width=0.4\linewidth]{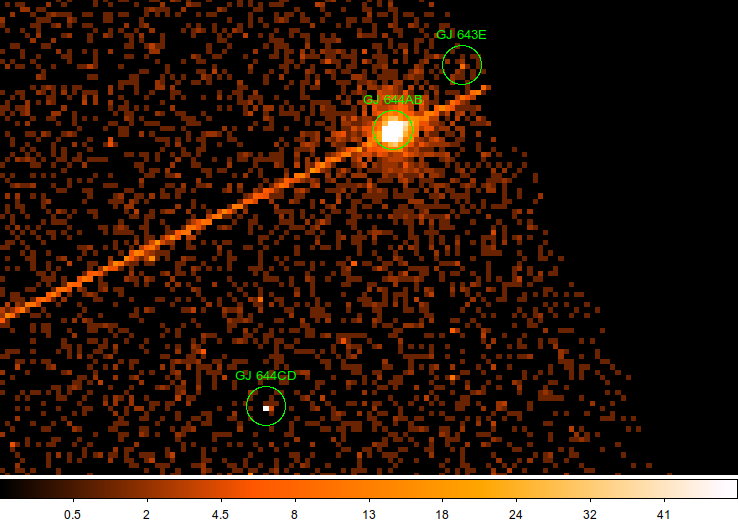}
    \caption{X-ray images of the GJ~644/GJ~643 system from ROSAT RASS (top left), ROSAT HRI (top right), \textit{XMM-Newton} EPIC (bottom left), and \textit{Chandra} ACIS (bottom right). The circles are around the stellar positions extrapolated back to the epoch of the respective observations and have a radius of $15^{\arcsec}$.}
    \label{fig:GJ643}
\end{figure*}

The majority of the ROSAT measurements were collected from the NEXXUS2\footnote{\href{https://hsweb.hs.uni-hamburg.de/projects/nexxus/nexxus.html}{https://hsweb.hs.uni-hamburg.de/projects/nexxus/nexxus.html}} catalog, the updated online version of the Nearby X-ray and XUV-emitting Stars data base \citep[NEXXUS;][]{NEXXUS}. Most stars in NEXXUS2 have been observed multiple times with various detectors: within the ROSAT All-Sky Survey (RASS) with the Position Sensitive Proportional Counter (PSPC), pointed observation with PSPC, and with the High Resolution Imager (HRI). We preferred observations with HRI over PSPC, and pointed observations were preferred over RASS. For multiple observations with the same detector, the one with the highest source detection likelihood (for PSPC) or signal-to-noise-ratio (for HRI) was taken. To convert the count rate into flux, we used the conversion factor listed in \cite{NEXXUS} ($6\times 10^{-12}\, \mathrm{erg \, cm}^{-2} \, \mathrm{ct}^{-1}$ for PSPC and $2.4\times 10^{-11}\, \mathrm{erg \, cm}^{-2} \, \mathrm{ct}^{-1}$ for HRI).

In NEXXUS2, measurements of non-resolved multiple systems are assigned to all components in the system. In this work, measurements of non-resolved systems were assigned either to the component that clearly lies closer to the X-ray source according to the 2MASS \textit{J}-band image or to the primary component if all components have similar distances to the X-ray source.

NEXXUS used quite large matching radii, $120^{\arcsec}$ for RASS, $60^{\arcsec}$ for pointed observation with PSPC and $30^{\arcsec}$ for HRI, which could lead to false matches. Therefore, we checked all matches in NEXXUS2 that have a positional offset between the X-ray source position and the star position that is larger than $15^{\arcsec}$. We inspected optical and 2MASS near-infrared images to check for other objects that might perhaps be a more likely counterpart of the X-ray source. 
In most cases,  the matches appear to be valid, but we identified three exceptional cases that are discussed in Appendix \ref{appendix:nexxus}.
 
We also found one case of a false match in NEXXUS2: GJ~643 (M3.5 at 6.5\;pc) is a component of a multiple system that also contains GJ~644AB (M3) and GJ~644CD (M7, better known as the star VB8). As is shown in Fig.~\ref{fig:GJ643}, the system is matched with RASS as well as HRI detections in NEXXUS2. Due to the low resolution of RASS, the system is not resolved and $L_\mathrm{X}$ from the resolved HRI detections were taken. However, no point source is present at the (proper-motion corrected) position of GJ~643 on the HRI image. NEXXUS2 reports a count rate of $0.00503\,\mathrm{s}^{-1}$ and an exposure time of 10705\;s for the star, which should give approximately 54 source counts. But within $15^{\arcsec}$ around the star's position there are only 19 counts, which is consistent with the background level. Therefore, this should be a false detection and GJ~643 is indeed not detected by ROSAT (fortunately, this is the only false detection we have found in NEXXUS2).

The GJ~643/644 system was also observed with \textit{XMM-Newton}. As can be seen in Fig.~\ref{fig:GJ643}, the point spread function of the bright source GJ~644AB extends even beyond the position of GJ~643, and indeed the 4XMM\_DR13 catalog does not list GJ~643 as an X-ray source. \cite{X_ray_catalog_10pc} have performed a customized data analysis for this star in which the authors determined the background level with regions at positions that have the same distance between GJ~643 and GJ~644AB to account for the contamination of the latter. In this way, they concluded that GJ~643 is indeed detected by \textit{XMM-Newton} with $L_\mathrm{X}$ on the order of $10^{26}\,\mathrm{erg\,s}^{-1}$.

However, there is an easier and more accurate way to determine the $L_\mathrm{X}$ of this star, which is to use the existing \textit{Chandra} ACIS data. As is shown on the bottom right panel of Fig.~\ref{fig:GJ643}, the system is well resolved by \textit{Chandra} and a faint source is clearly seen at the position of GJ~643. By using the \texttt{srcflux} command described in Sect. \ref{sec:chandra}, we determined the X-ray flux $F_\mathrm{X}=9.41\times 10^{-13}\, \mathrm{erg\,cm}^{-2}\,\mathrm{s}^{-1}$ in the 0.5$-$7~keV band for this star, corresponding to a $L_\mathrm{X}$ of $6.93\times 10^{25}\,\mathrm{erg\,s}^{-1}$ in the ROSAT band.

\subsubsection{2RXS / 2RXP}

As a complement to NEXXUS2, we also cross-matched our samples with the Second ROSAT all-sky survey source catalog \citep[2RXS;][]{2RXS} within $15^{\arcsec}$. The background-corrected count rate was converted to flux with the conversion factor reported in 2RXS ($1.08\times 10^{-11}\, \mathrm{erg \, cm}^{-2} \, \mathrm{ct}^{-1}$). There are altogether 11 M stars, 3 K stars, and 1 G stars matched additionally in 2RXS.

Finally, we also cross-matched our samples with the Second ROSAT Source Catalog of Pointed Observations \citep[2RXP;][]{2RXP}, which contains ROSAT pointed observations with PSPC between 1990 and 1997. 
This yielded two additional M stars.
The count rate was finally converted to flux with the same conversion factor as in 2RXS.

\subsection{eROSITA}

Most of the eROSITA X-ray luminosities used in this work were taken from \cite{eROSITA_M_dwarf_sample}, who constructed an M star sample with cross-correlation of the SUPERBLINK proper motion catalog of nearby M dwarfs containing M stars within 100~pc \citep{LG11} and \textit{Gaia} DR2. X-ray fluxes of the stars in the paper were collected from the first public data release of eROSITA (eRASS1) for 690 stars. All 36 M dwarfs that have distances smaller than 10~pc in the paper are contained in the M star sample of this work.

\cite{X_ray_catalog_10pc} have also constructed a complete X-ray sample for nearby M stars with SpT $\le$ M4 based on data collected from the merged first four all-sky surveys (eRASS1 to eRASS4) of eROSITA, ROSAT and \textit{XMM-Newton}, with only one star, GJ~745A (M2 at 8.8~pc), not detected.  Their paper contains eROSITA data of seven M stars that were not detected in eRASS1, and we have extracted their values from Fig. 2 in the paper.  
One of the stars (GJ~367, M1 at 9.4~pc) was analyzed separately in detail \citep{GJ367}, and we adopted its X-ray flux from this individual analysis.

The data from the first public data release of the eROSITA telescope \citep[eRASS1;][]{erass1,Freund2024} is considered complementary to the data found previously. Crossmatching with the eRASS1 public catalog within a matching radius of $15^{\arcsec}$ gives two further detections for M stars, four for K stars, and five for G stars.

\subsection{Determinations of upper limits for X-ray undetected stars}\label{sec:upper_limit}

For the majority of the X-ray undetected stars, we either used the High-Energy Lightcurve Generator (HILIGT) upper limit server\footnote{\href{http://xmmuls.esac.esa.int/upperlimitserver/}{http://xmmuls.esac.esa.int/upperlimitserver/}} \citep{HILIGT} to determine their $L_\mathrm{X}$ upper limits from ROSAT and/or \textit{XMM-Newton} observations, or the data products of eRASS1 from eROSITA observations \citep{erass_upper_limit}. HILIGT calculated the X-ray source count and background count at the position of the star within the 0.2$-$2\;keV energy band, while eRASS1 calculated them within 0.2$-$2.3\;keV. These were inserted into the Python function \texttt{astropy.stats.poisson\_conf\_interval} to calculate the upper limit photon count at a 90\% confidence level using a Bayesian method \citep{poisson_conf_interval}. Finally, the photon count rates were converted to $L_\mathrm{X}$ in the ROSAT band with the WEBPIMMS tool\footnote{\href{https://heasarc.gsfc.nasa.gov/cgi-bin/Tools/w3pimms/w3pimms.pl}{https://heasarc.gsfc.nasa.gov/cgi-bin/Tools/w3pimms/w3pimms.pl}} for the HILIGT upper limits. For the eRASS1 results, the conversion factor in \cite{X_ray_catalog_10pc} was used for M stars ($8.78\times 10^{-13}\, \mathrm{erg \, cm}^{-2} \, \mathrm{ct}^{-1}$) to convert the count rate to ROSAT flux, and the conversion factor in \cite{erass1} was used for G/K stars ($9.31\times 10^{-13}\, \mathrm{erg \, cm}^{-2} \, \mathrm{ct}^{-1}$) to convert it to flux in the 0.2$-$2.3~keV band.

For some stars that have other bright X-ray sources in the field of view, however, HILIGT can falsely assign the other detected sources as matches and give out a detection. In these cases, we calculated the upper limits with RASS images, which is described in detail in Appendix \ref{app: upper_limit}.

For some stars, we found upper limits in the literature. Upper limits calculated from RASS observations of 14~M stars were taken from \cite{10pc_XUV}, and the upper limit of GJ~745A based on a \textit{XMM-Newton} pointed observation was taken from \cite{X_ray_catalog_10pc}.

\subsection{Conversion between energy bands}\label{sec:conversion}

As the X-ray luminosities (and upper limits) measured by different instruments are within different energy bands, they need to be converted to the same energy band for comparison, for which the ROSAT energy band (0.1--2.4\;keV) was chosen. The conversion factor between energy bands were calculated with WebPIMMS for a single-temperature APEC plasma spectrum model with solar abundance. 
Since the interstellar column densities for nearby ($d \leq  25$\;pc) stars are generally very small \citep[$N_\text{H} \la 10^{18}\, \mathrm{cm}^{-2}$, ][]{NH_d},
extinction in the X-ray band is very small and can be neglected. 

\begin{figure}[htbp]
    \centering
    \includegraphics[width=7.5cm]{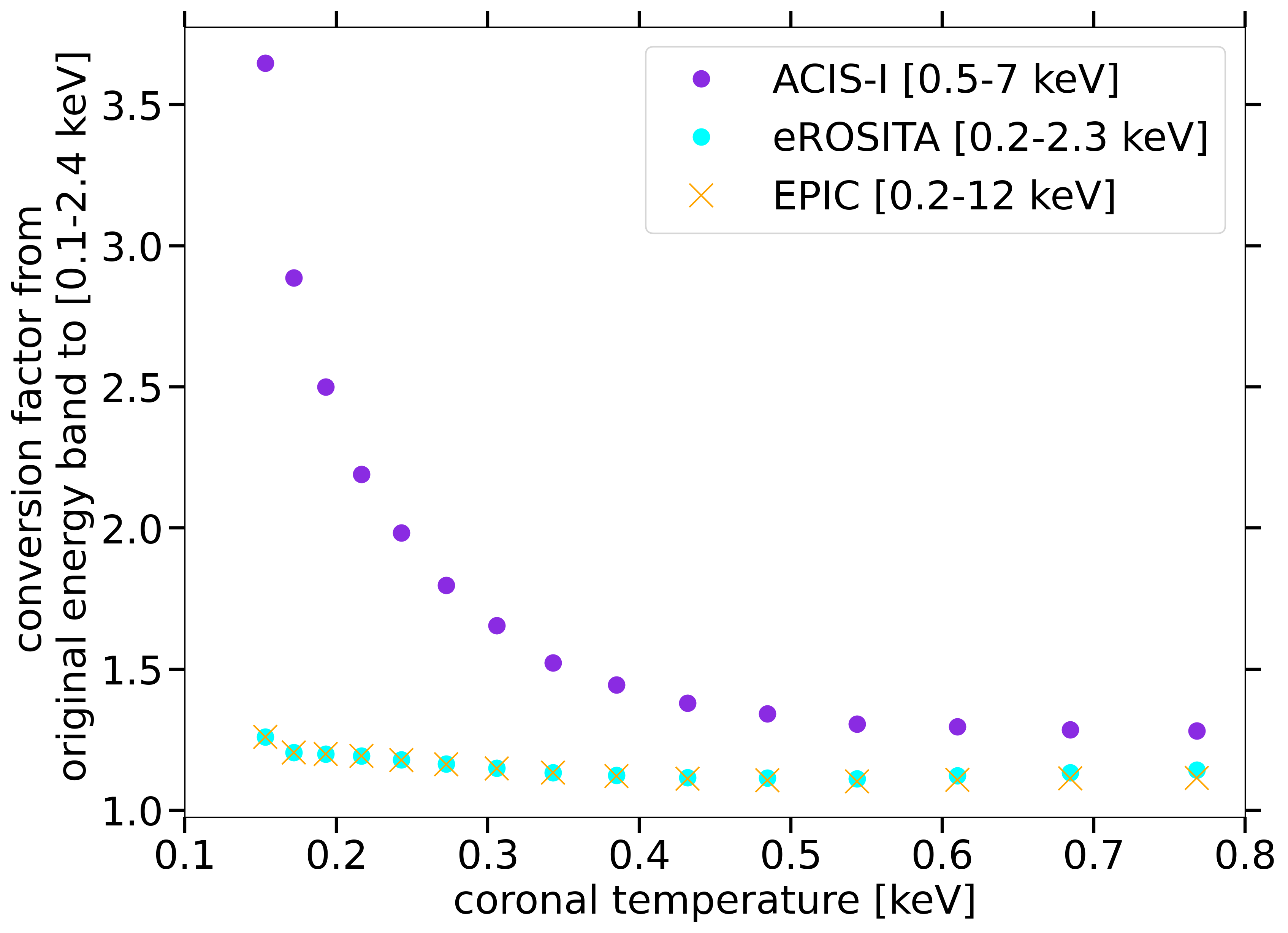}
    \caption{Conversion factor from the energy bands of \textit{Chandra} ACIS-I, eROSITA, and \textit{XMM-Newton} EPIC to the ROSAT energy band with the APEC model of solar abundance and no extinction plotted against coronal temperatures. }
    \label{fig:conversion_factor}
\end{figure}

For 31 M stars, 8 K stars, and 6 G stars with coronal temperature determinations in the literature, we either adopted their values directly (see Tables~\ref{tab:kT} and~\ref{tab:kT2}) or calculated an averaged temperature for multitemperature models (see Table~\ref{tab:kT3} and the related discussion). For stars that do not have literature measurements of their coronal temperatures, $kT=0.5\,\mathrm{keV}$ was used as an approximation for the M stars and $kT=0.3\,\mathrm{keV}$ was used for G/K stars.

Figure~\ref{fig:conversion_factor} shows the conversion factor between different energy bands calculated with WebPIMMS, plotted against coronal temperatures. Literature measurements of the coronal temperature of M stars in our sample typically range from $0.16\,\mathrm{keV}$ to $0.77\,\mathrm{keV}$ (the M4.5 star GJ~799A with a coronal temperature over $1\,\mathrm{keV}$ is unusual and is also discussed in Sect.~\ref{sec:KM_M}). For G/K stars, the temperature typically ranges from $0.15\,\mathrm{keV}$ to $0.55\,\mathrm{keV}$ \citep[the two G stars with low coronal temperature below $0.1\,\mathrm{keV}$, GJ~672 and GJ~882, are Maunder Minima candidates that are probably during their low-activity period;][]{2009A&A...508.1417P, 2012IAUS..286..335S}. Within this typical temperature range, the conversion factors between the different energy bands vary between 1.3 and 1.1 for
\textit{XMM-Newton} EPIC and eROSITA; 
for \textit{Chandra} ACIS-I, the conversion factors are in the range of $\approx 1.3 ... 1.5$ for plasma temperatures $\ge 0.3$\;keV, but rise to  values of up to 3.7 for very soft spectra.

For stars without a plasma temperature measurement, the assumption of a constant ``typical'' plasma temperature and a corresponding conversion factor per SpT bin introduces some degree of uncertainty into the X-ray luminosity estimate,
since the plasma temperatures are often found to be correlated to X-ray surface fluxes \citep[see e.g.][]{T_Fx_relation}.
However, Fig.~2 suggests that this uncertainty should be
less than a factor of two in the vast majority of stars.

It is well established that the X-ray luminosities of stars show intrinsic variability, typically (at least) by a factor of two to three; for example,~over a stellar activity cycle \citep[see e.g.][for the activity cycles of the Sun and $\alpha$ Cen A/B]{alpha_cen_Lbol}.  The uncertainty of our X-ray luminosity estimates resulting from the assumption of a constant coronal temperature for each spectral class (for stars for which no plasma temperature determination are available in the literature) is comparable with the intrinsic stellar variability.

Another source of uncertainties in the determination of X-ray luminosities can be stellar flares.
During particularly strong X-ray flares, 
the X-ray luminosity can rise by considerable factors within minutes to hours, followed by a slower (often approximately exponential) decay phase back to the ``quiescent'' value.
Long X-ray observations of bright X-ray sources (yielding high-signal-to-noise light curves) can allow for a detailed analysis of individual flare events 
\citep{review_flares},
and a meaningful discrimination between flaring and quiescent phases.
However, a reliable distinction of these phases is often not possible, as the apparently quiescent emission may in fact contain (or even be dominated by) numerous low-amplitude flares, and the minimum amplitude of a ``detectable'' flare depends strongly on the count rate of the source. 

In this work, we necessarily have to deal with an inhomogeneous collection of various  X-ray observations with very different exposure times, and the X-ray detections of our target stars have a wide range of different signal-to-noise ratios. This prevents a homogeneous and consistent analysis of flares in the light curves of individual stars.
We have therefore used the X-ray luminosity determined as a temporal mean value over the full duration of the observation to characterize the activity level.

\section{Properties of the final samples}\label{sec:discussion}

Many stars in our samples have been detected multiple times with different instruments. In these cases, $L_\mathrm{X}$ measured by \textit{Chandra} or \textit{XMM-Newton} was used for the X-ray analysis, if available; otherwise, measurements from eROSITA were used, or the last priority, ROSAT.

\subsection{X-ray detection completeness}\label{sec:completenss}

\begin{figure}[htb]
    \centering
    \includegraphics[width=7.8cm]{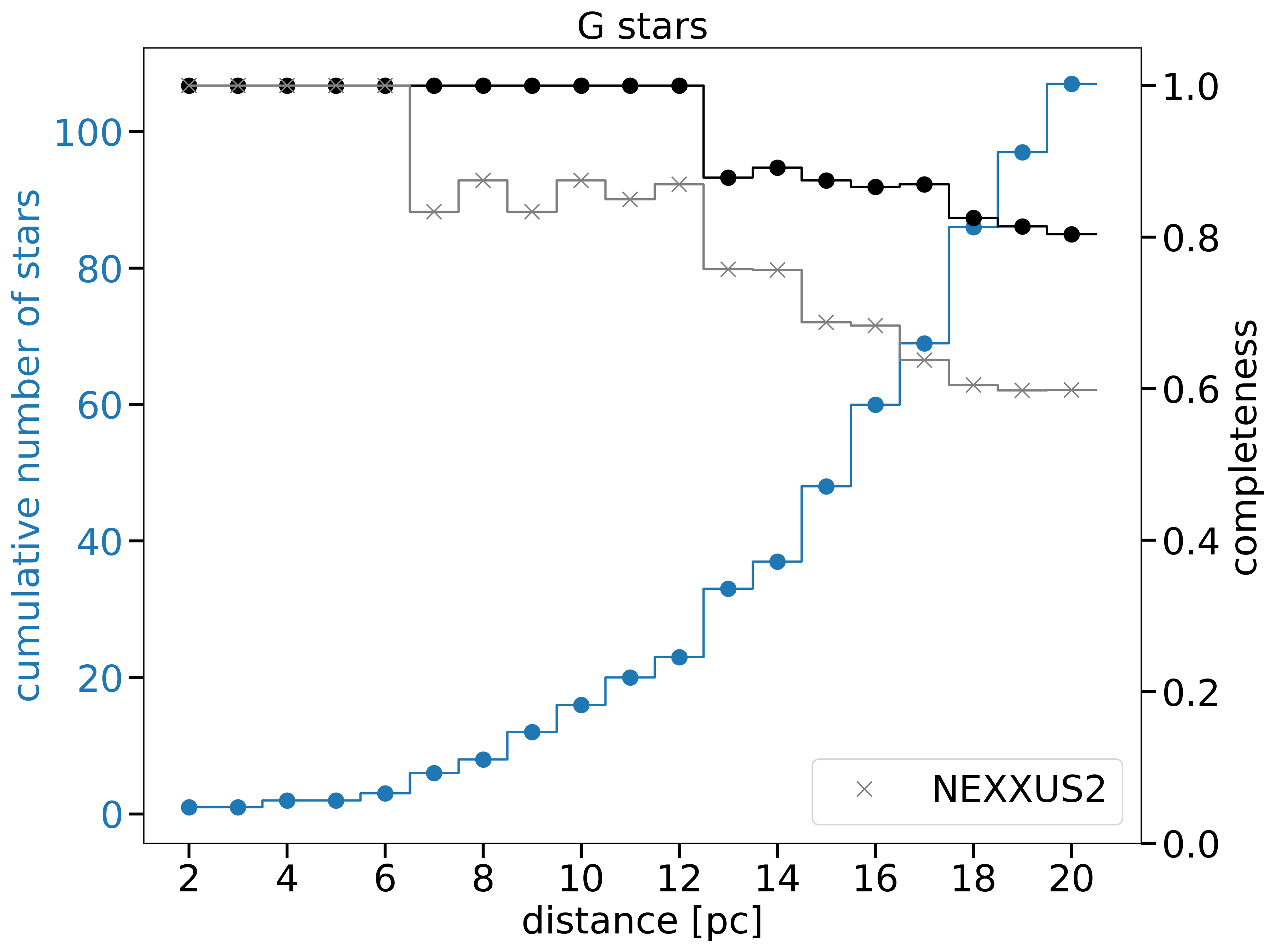}
    \includegraphics[width=7.8cm]{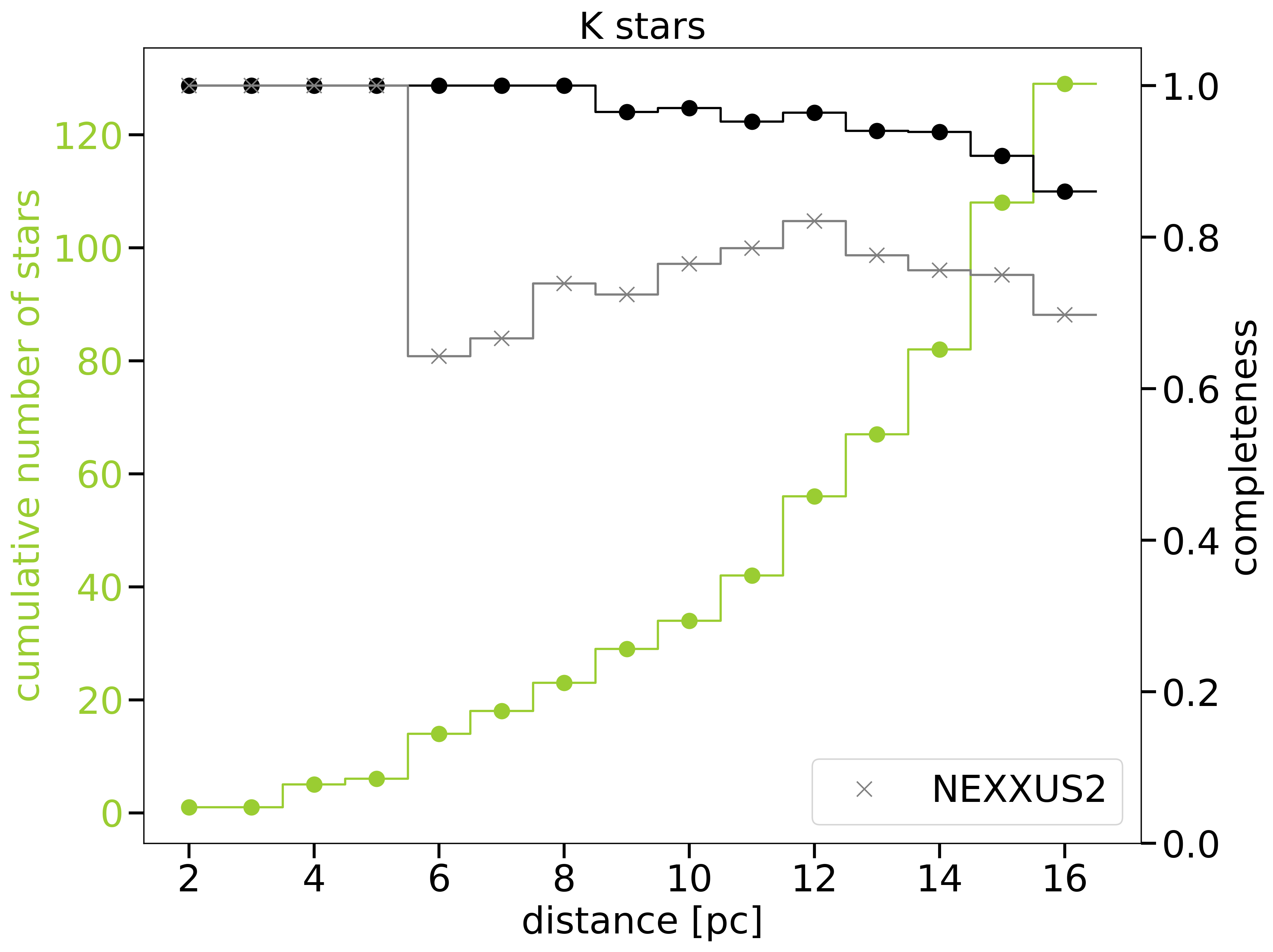}
    \includegraphics[width=7.8cm]{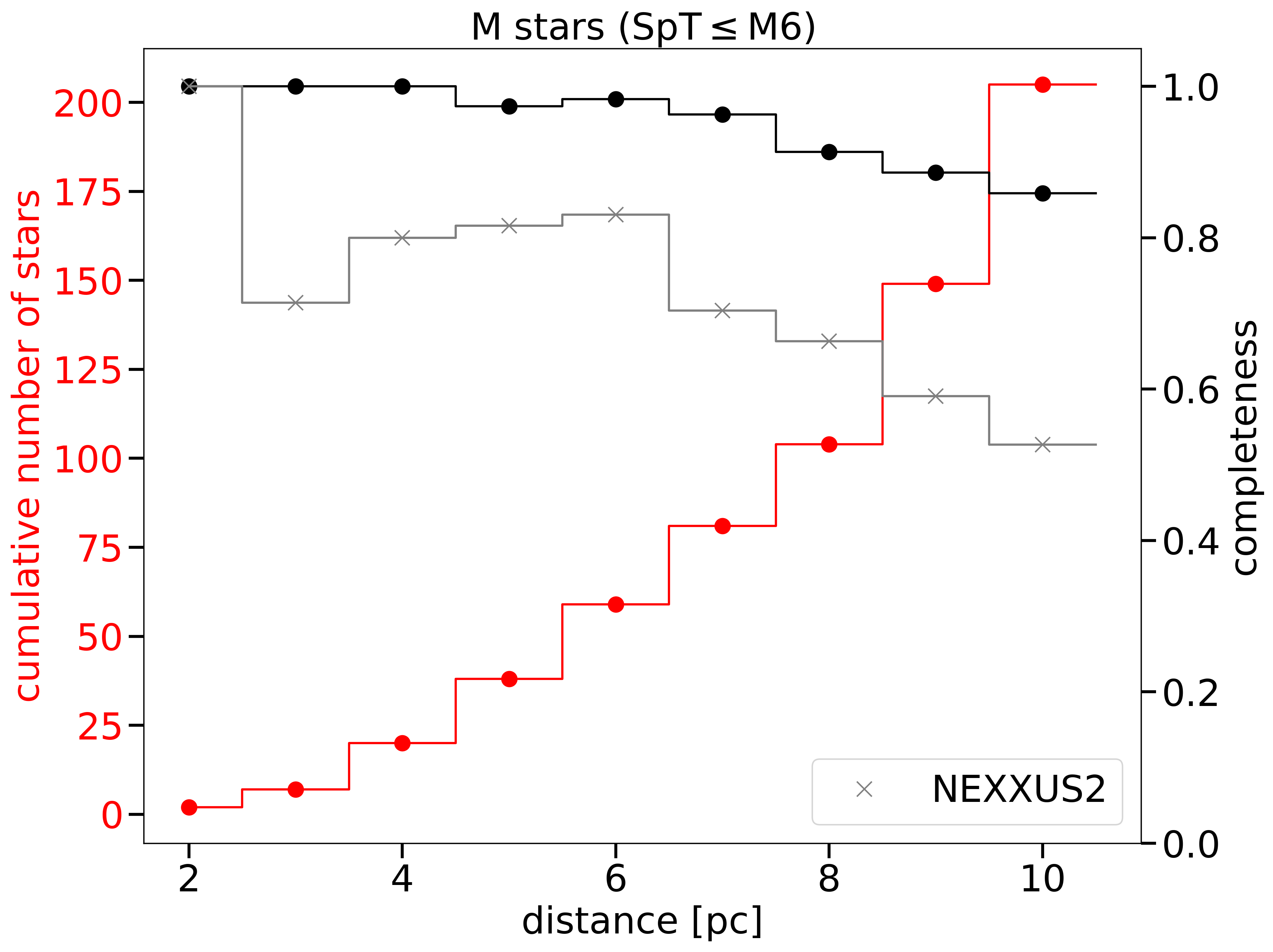}
    \caption{Cumulative number of stars and X-ray detection completeness plotted against distance for different SpTs. Values of this work are shown in circles; those of NEXXUS2 are shown in crosses.} 
    \label{fig:star_number}
\end{figure}

Figure~\ref{fig:star_number} shows the cumulative number of stars of different SpT and X-ray detection completeness of this work plotted against distance. There are 205 M stars, 129 K stars, and 107 G stars in the sample with distances of up to 10~pc, 16~pc, and 20~pc, respectively. The X-ray detection completeness is 85\% for M stars, 86\% for K stars, and 80\% for G stars. An X-ray detection completeness of over 80\% makes sure that all samples are complete enough to perform a meaningful comparison of the X-ray luminosity distribution functions. It also ensures that the K star and G star samples have comparable sizes with the M star sample and can be compared with each other.

Compared with previous published X-ray samples of nearby stars, the samples of this work represent substantial progress. 
Figure~\ref{fig:star_number} shows the X-ray detection completeness of NEXXUS2. One can see that due to the improvements of the stellar census in the solar neighborhood in the past 20 years, NEXXUS2 is no longer complete within its reported completeness limit \citep[6~pc for M stars, 12~pc for K stars and 14~pc for G stars;][]{NEXXUS}.  
The sample size of this work is also more than twice the sample size of NEXXUS.

\subsection{X-ray luminosities and stellar distances }\label{sec:Lx}

\begin{figure}[htbp]
    \centering
    \includegraphics[width=7.8cm]{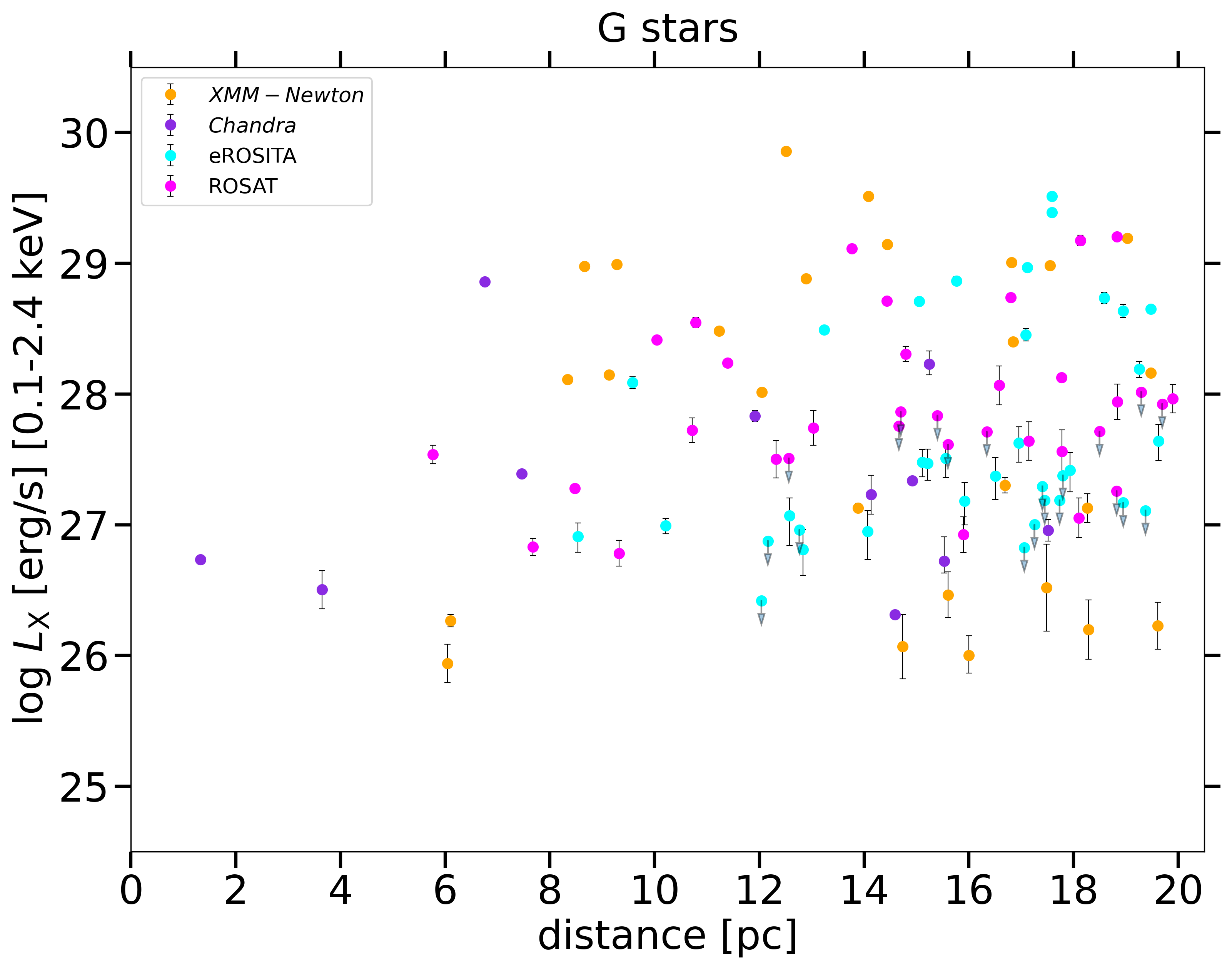}
    \includegraphics[width=7.8cm]{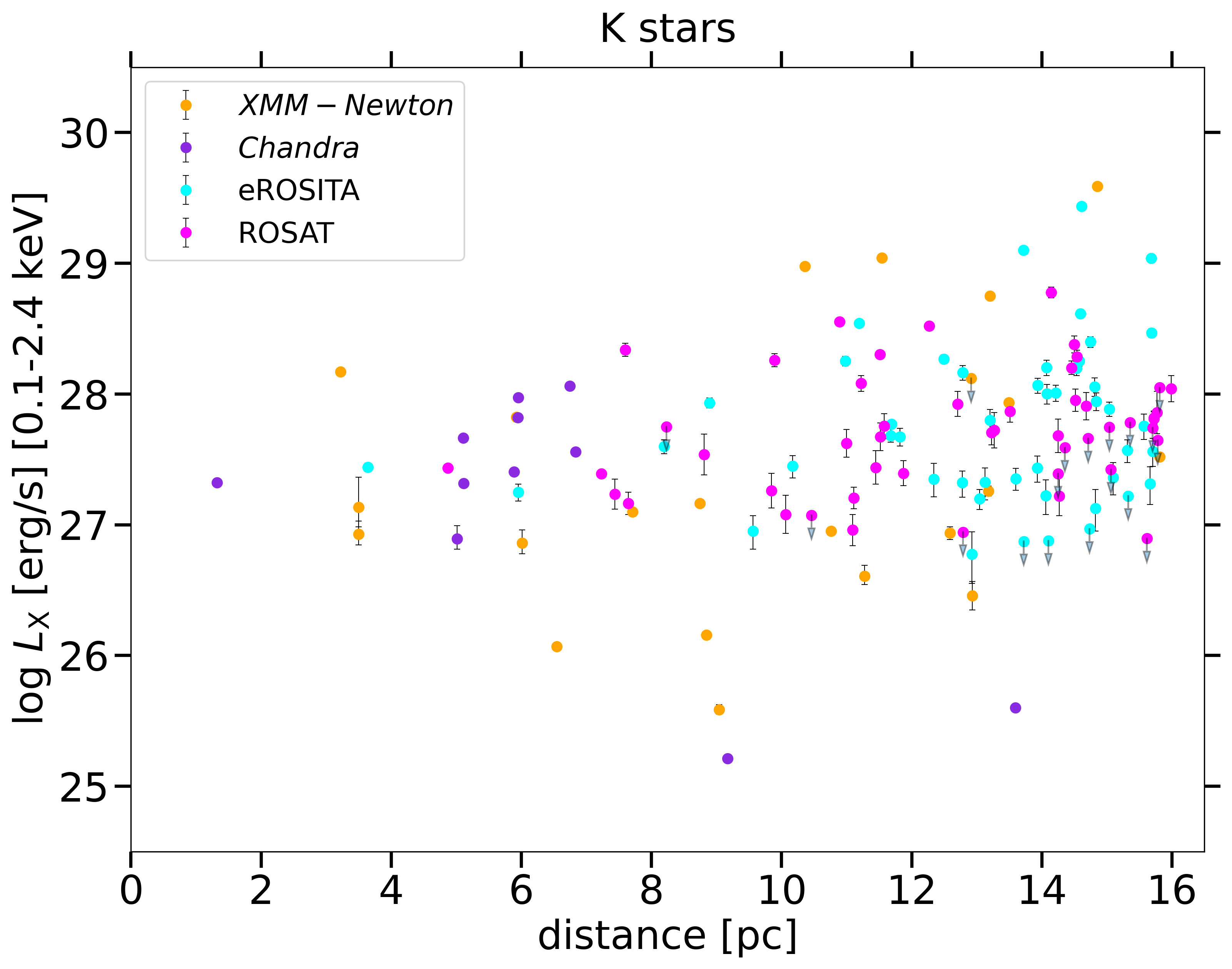}
    \label{fig:completeness_M6_fin}
    \includegraphics[width=7.8cm]{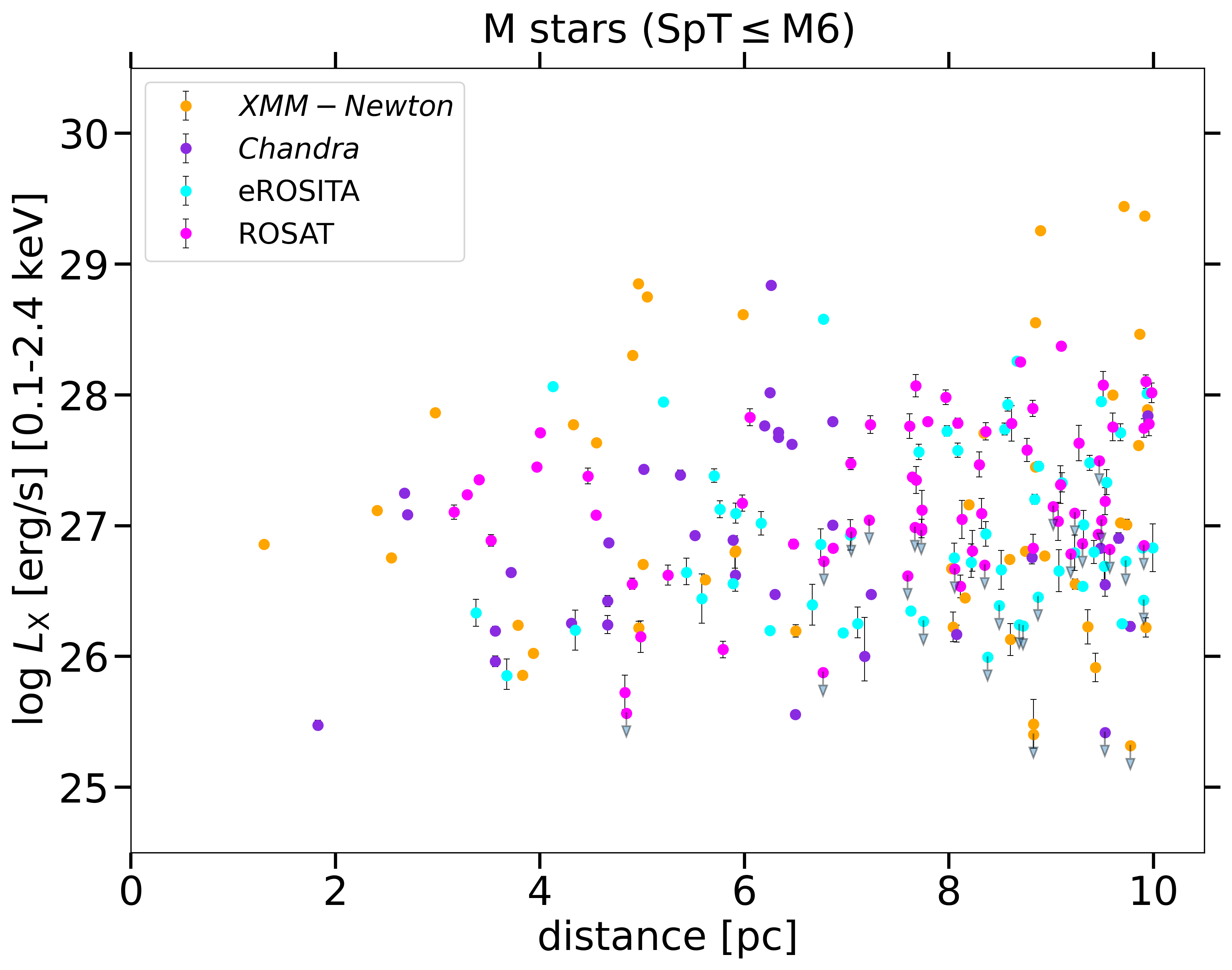}
    \caption{X-ray luminosity vs. distance for the G-, K-, and M-type stars in our samples. Different colors represent measurements with different instruments. Downward-pointing arrows indicate upper limits.}
    \label{fig:completeness_all}
\end{figure}

Figure~\ref{fig:completeness_all} shows X-ray luminosity plotted against distance for stars of different SpTs from the sample of this work. One can see that for all SpTs, $L_\mathrm{X}$ is not dependent on distance and the lower boundary is nearly flat, which indicates that our samples are unbiased.

In the following analysis, 46 M stars have $L_\mathrm{X}$ or upper limits measured by \textit{XMM-Newton}, 35 by \textit{Chandra}, 55 by eROSITA, and 69 by ROSAT. Twenty-two K stars have $L_\mathrm{X}$ or upper limits measured by \textit{XMM-Newton}, 11 by \textit{Chandra}, 47 by eROSITA, and 49 by ROSAT. Twenty-six G stars have $L_\mathrm{X}$ or upper limits measured by \textit{XMM-Newton}, 11 by \textit{Chandra}, 36 by eROSITA, and 34 by ROSAT.

\subsection{Comparison with NEXXUS}

\begin{figure}[htbp]
    \centering
    \includegraphics[width=8cm]{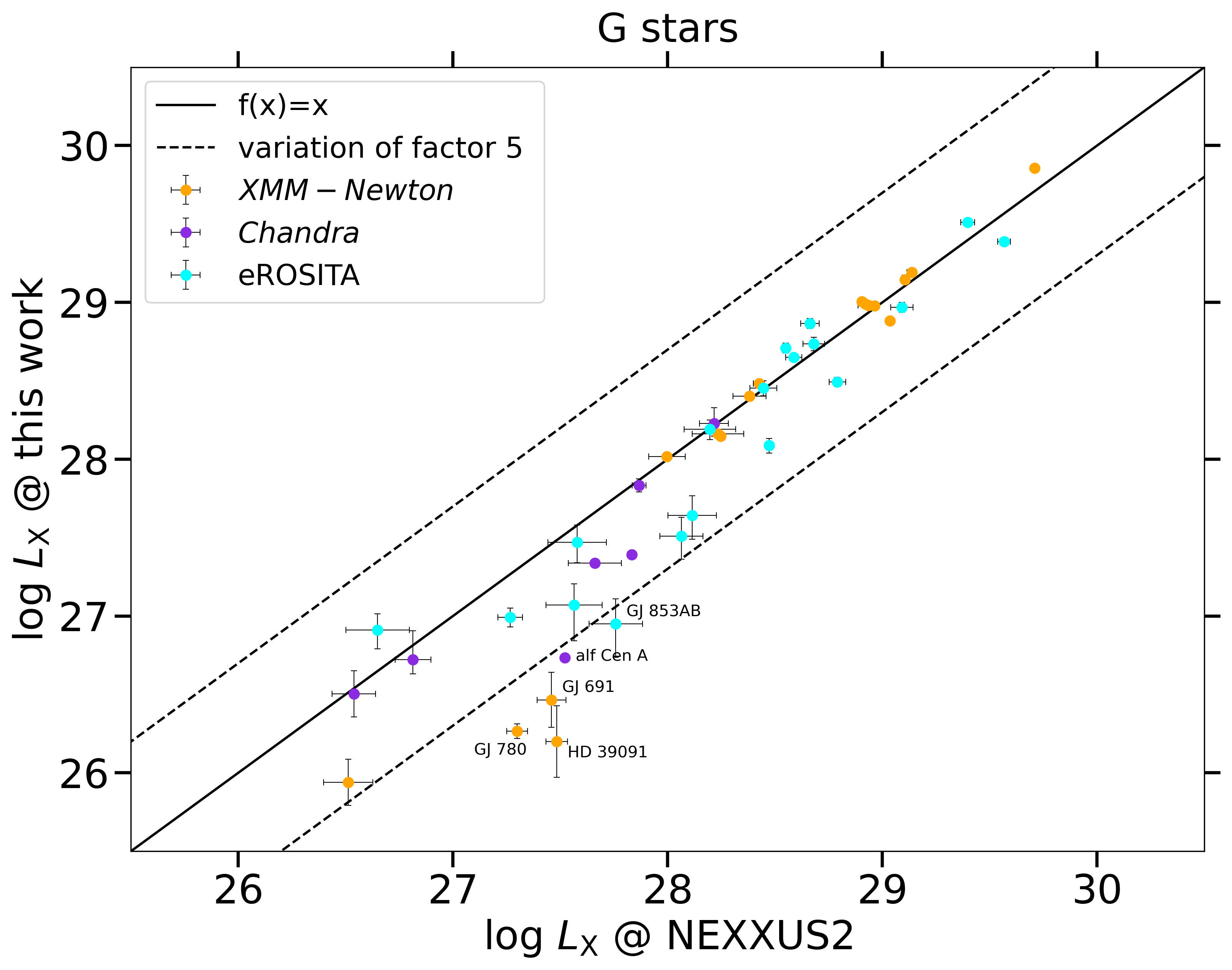}
    \includegraphics[width=8cm]{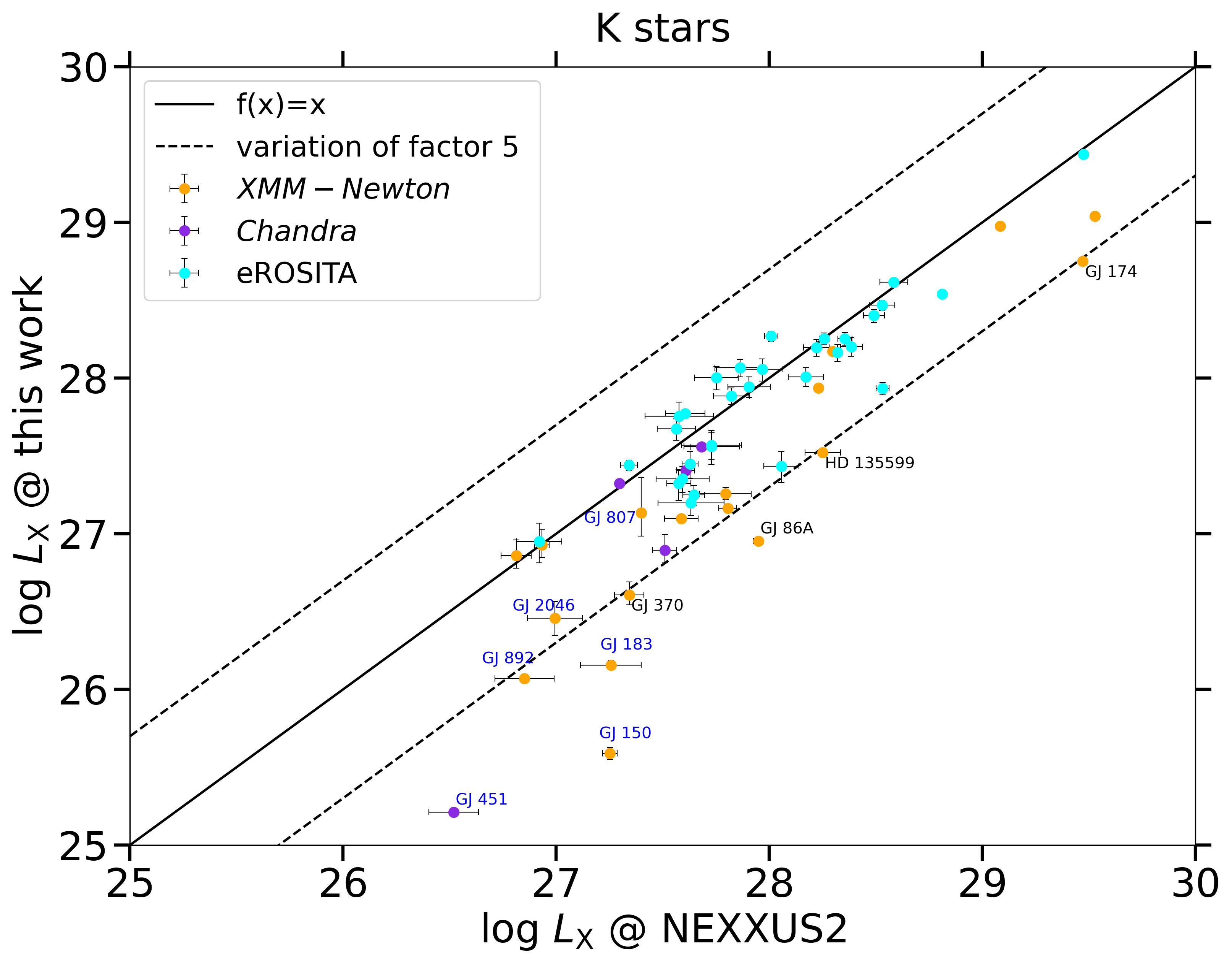}
    \includegraphics[width=8cm]{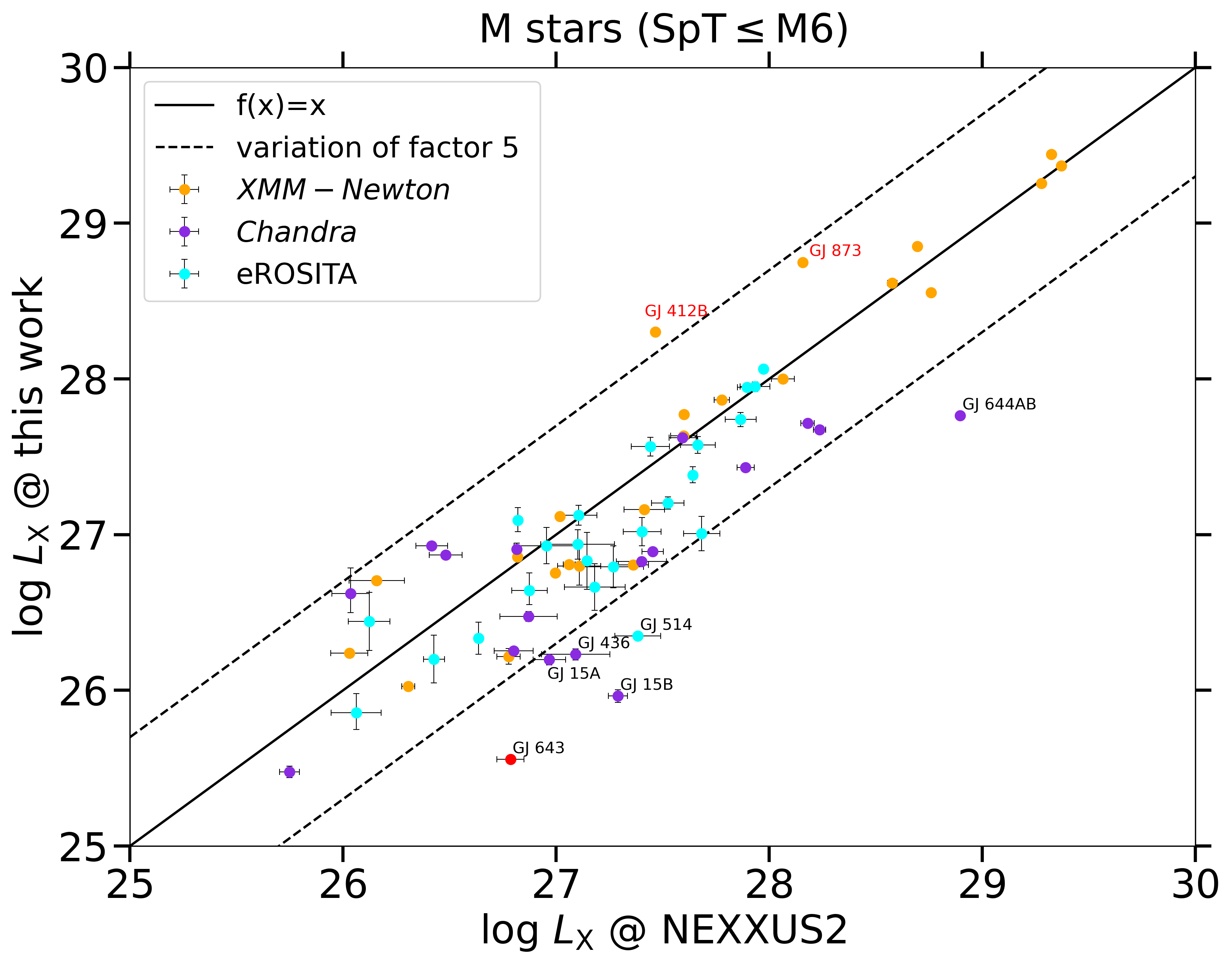}
    \caption{X-ray luminosities of the stars that were measured by other instruments than ROSAT plotted against $L_\mathrm{X}$ in NEXXUS2. Two M stars with $\log(L_\mathrm{X}/L_\mathrm{bol})>-3$ are annotated with red text and six K stars with $\log(L_\mathrm{X}/L_\mathrm{bol})<-6.5$ are annotated with blue text. GJ~643, with the wrongly assigned value in NEXXUS2, is also shown.} 
    \label{fig:nexxus_comparison}
\end{figure}

Figure~\ref{fig:nexxus_comparison} shows the $L_\mathrm{X}$ of the stars from NEXXUS2 plotted against the values from other instruments in our samples. 
For most stars, the X-ray luminosities agree within a factor of a few.
The points below the diagonal show how measurements with higher sensitivity and better angular resolution increase the deepness of our samples compared to NEXXUS2.

For a few stars, we find X-ray luminosities that are considerably smaller than the values listed in NEXXUS2 (less than one fifth the size of them), especially at the low luminosity end. A possible explanation for these cases is that these (relatively X-ray faint) stars could only be detected by ROSAT because they showed a stellar flare, or a period of high activity with enhanced X-ray luminosities, during the ROSAT observation, while the other instruments with higher sensitivity were able to detect these stars during more quiescent periods.

\subsection{Distributions of $L_\mathrm{X}/L_\mathrm{bol}$}\label{sec:cumulative_distribution}

\begin{figure}[htbp]
    \centering
    \includegraphics[width=8cm]{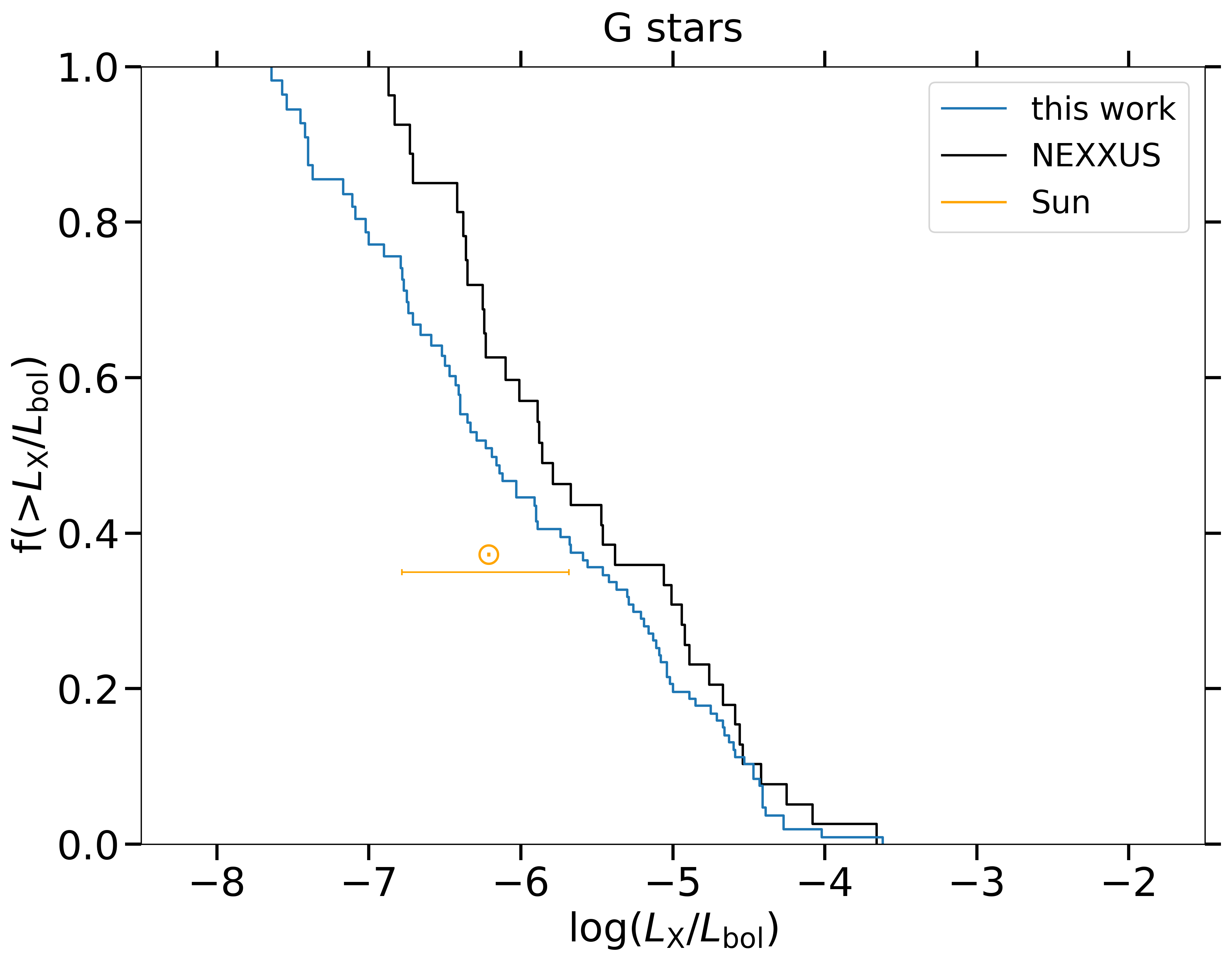}
    \includegraphics[width=8cm]{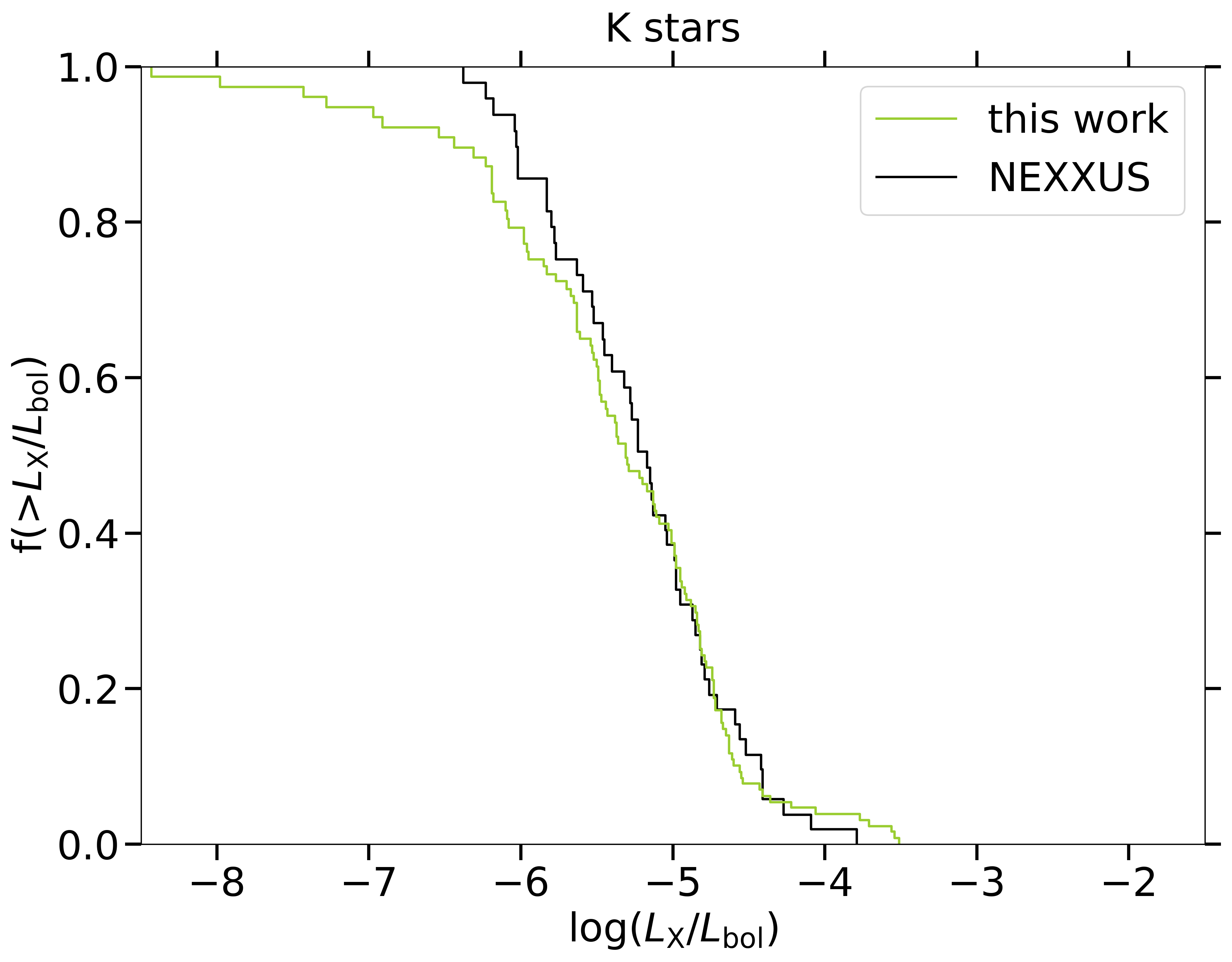}
    \includegraphics[width=8cm]{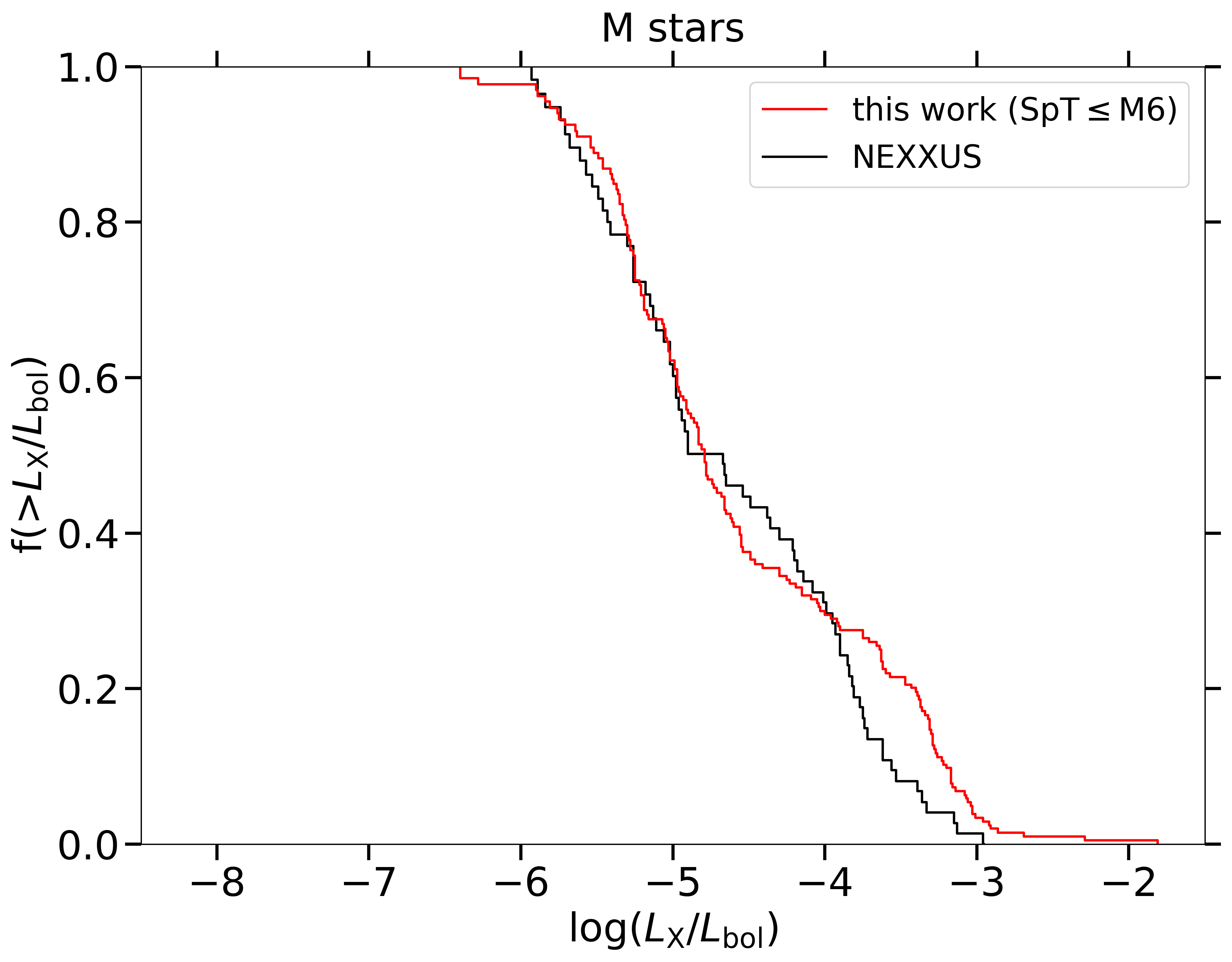}
    \caption{Kaplan-Meier estimator for the distributions of $L_\mathrm{X}/L_\mathrm{bol}$ of G, K, and M stars in the samples of this work, compared with distributions of NEXXUS. The solar value extrapolated to a full solar cycle \citep{Sun_Lx} is also shown in the plot for G stars.}
    \label{fig:cumulative_all}
\end{figure}

To calculate the distribution function of $L_\mathrm{X}/L_\mathrm{bol}$ in our samples that include upper limits, we used the software Astronomy SURVival analysis \citep[ASURV;][]{asurv}. ASURV computes the Kaplan-Meier estimator, which mathematically distributes the upper limits within a certain dataset to the nearest measurements and calculates the distribution function of the dataset \citep{Kaplan_Meier, Kaplan_Meier2}. The Kaplan-Meier estimator for the distributions of $L_\mathrm{X}/L_\mathrm{bol}$ of stars in the samples of this work are shown in Fig.~\ref{fig:cumulative_all}, in which they are also compared with the distributions of stars in NEXXUS \citep{NEXXUS}.

\subsubsection{G stars}

The top panel of Fig.~\ref{fig:cumulative_all} shows the comparison of the Kaplan-Meier estimator for the distributions of $\log(L_\mathrm{X}/L_\mathrm{bol})$ of G  stars of this work with the results of NEXXUS. The solar value, which ranges between $-5.68$ and $-6.78$ over one solar cycle \citep{Sun_Lx}, is also plotted. The distribution of G stars from this work extends to considerably lower $L_\mathrm{X}/L_\mathrm{bol}$ values than the one from NEXXUS (mostly because the measurements from \textit{Chandra} and \textit{XMM-Newton} can detect much fainter sources than ROSAT), and the sample size of this work is about 2.5 times the sample size of NEXXUS. According to the result of this work, the Sun has a $\log(L_\mathrm{X}/L_\mathrm{bol})$ value very close to the median value for the G-type stars; in other words, the Sun is a typical G star.

\subsubsection{K stars}

The K star distribution of this work has a low $L_\mathrm{X}/L_\mathrm{bol}$ tail that the NEXXUS distribution does not have, contributed by stars with $\log (L_\mathrm{X}/L_\mathrm{bol})<-6.5$ (annotated with blue text in the middle panel of Fig.~\ref{fig:nexxus_comparison}). There are four K stars that have very low $\log(L_\mathrm{X}/L_\mathrm{bol})<-7$: GJ~150 (K0 subgiant at 9.0~pc), GJ~451 (K1 dwarf at 9.2~pc), GJ~615 (K3 dwarf at 13.6~pc), and GJ~807 (K0 subgiant at 14.3~pc). 
The two K type subgiants have $L_\mathrm{bol}$ comparable with main sequence F and A stars, respectively, which might be the main reason for their low $L_\mathrm{X}/L_\mathrm{bol}$. GJ~615, on the contrary, is an X-ray faint star with $L_\mathrm{X} = 3.99\times 10^{25}\,\mathrm{erg\,s}^{-1}$ that is detected only with \textit{Chandra}.

\subsubsection{M stars}\label{sec:KM_M}

The bottom panel of Fig.~\ref{fig:cumulative_all} shows the Kaplan-Meier estimator for the distributions of $\log(L_\mathrm{X}/L_\mathrm{bol})$ of M stars. 
The distribution shows a high activity tail that NEXXUS does not have, contributed by stars with $\log(L_\mathrm{X}/L_\mathrm{bol})>-3$. There are altogether seven such M stars, of which only two 
were also included in NEXXUS; they are annotated with red text in the bottom panel of Fig.~\ref{fig:nexxus_comparison}. Both of them are known flare stars. One star, GJ~799A = AT Mic at 9.9~pc, shows an extremely high X-ray activity level of $\log(L_\mathrm{X}/L_\mathrm{bol})=-1.81$. It is a visual binary system consisting of two frequently flaring M4.5 stars with an angular separation of $2.1\arcsec$, and was observed by \textit{XMM-Newton} \citep{GJ799}. A flare occurred during that observation, during which $L_\mathrm{X}$ was 1.7 times the quiescence level.

\section{Activity of M-type stars and comparison to G- and K-type stars}
\label{sec:LxLbol-GKM}

\begin{figure}[htbp]  
    \centering
    \includegraphics[width=8.0cm]{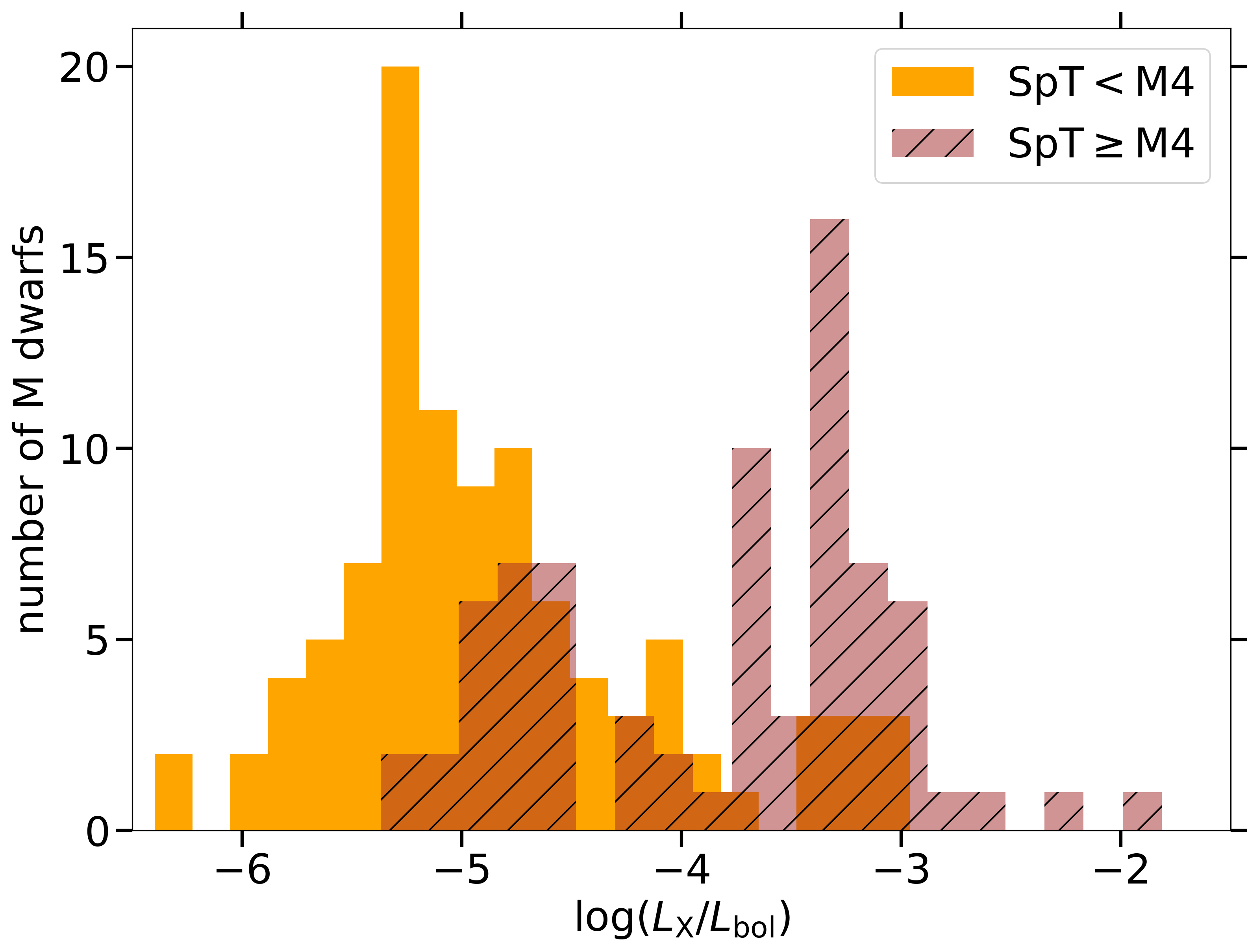}  \\
    \includegraphics[width=8.0cm]{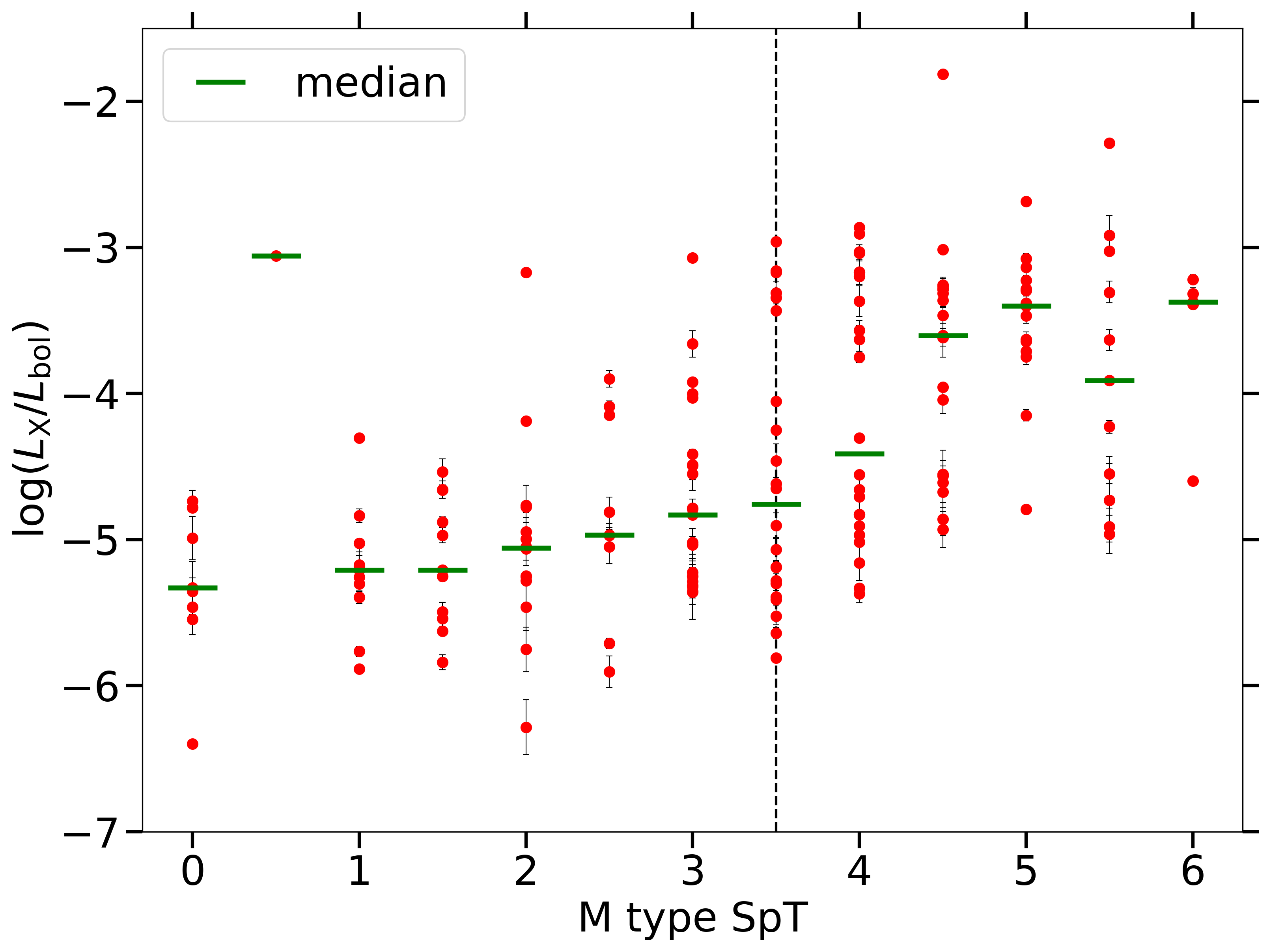} \\
    \includegraphics[width=8.0cm]{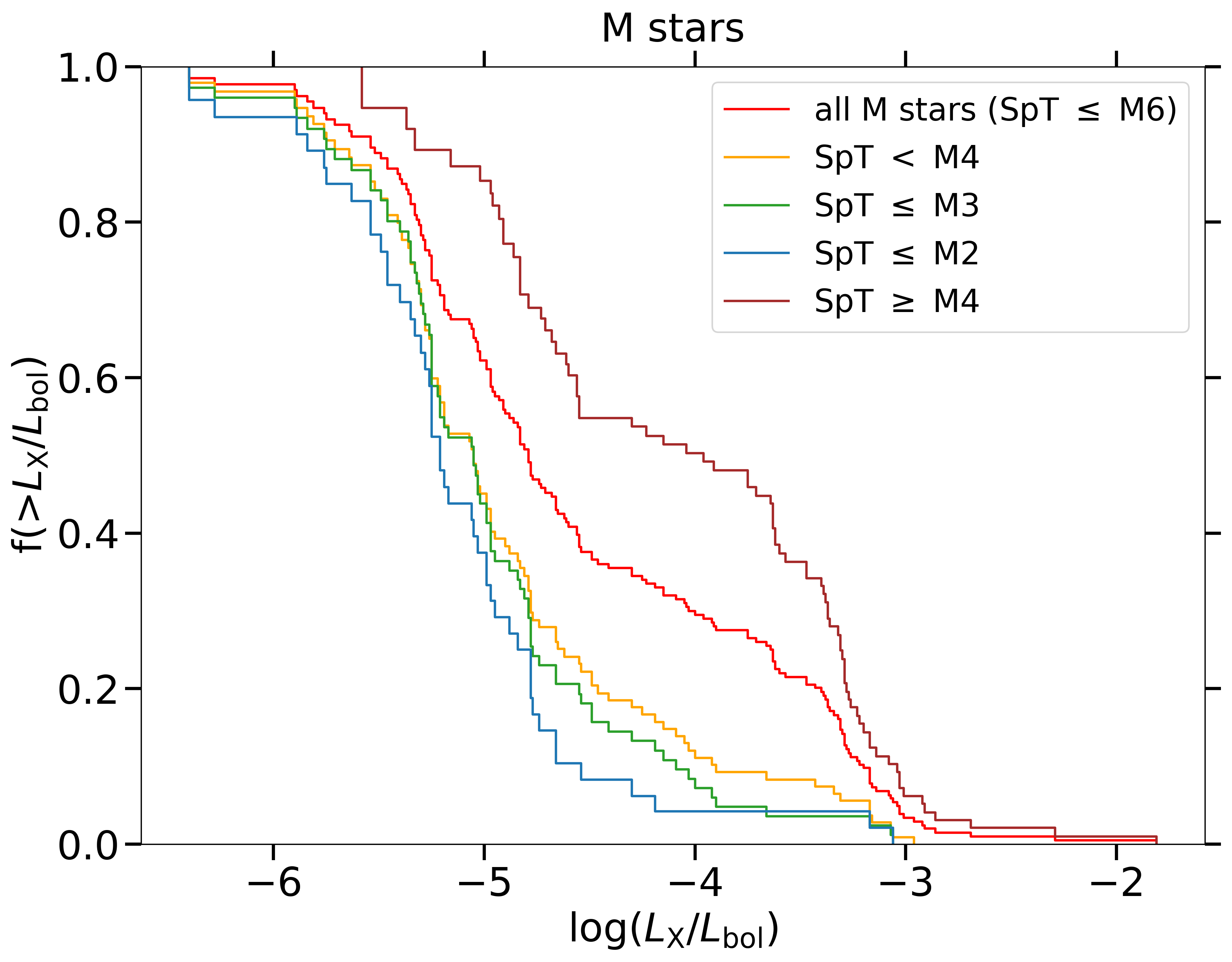}
    \caption{The difference in X-ray activity between early and late M-type stars shown from different perspectives. \textit{Upper panel:} Histogram of $\log(L_\mathrm{X}/L_\mathrm{bol})$ of all X-ray detected M stars in our sample, separated into two subsamples with SpT $\le$\,M4 and $>$\,M4. \textit{Middle panel:} $\log(L_\mathrm{X}/L_\mathrm{bol})$ of all X-ray detected M stars are plotted against spectral subtype. The median value of each subtype is shown with a horizontal green line. A vertical dashed line is drawn at SpT M3.5.
    \textit{Lower panel:} Kaplan-Meier estimator for the distributions of 
     $\log(L_\mathrm{X}/L_\mathrm{bol})$ for
 all M stars in our sample and of different subsamples based on their SpT.}
    \label{fig:histogram_M}
\end{figure}

\subsection{X-ray activity for early and late M-type stars}

The histogram of $\log(L_\mathrm{X}/L_\mathrm{bol})$ for the X-ray detected M-type stars (Fig.~\ref{fig:histogram_M}, top) shows a bimodal distribution, with one peak at about $-5$ and another peak lying between $-4$ and $-3$. The middle part of Fig.~\ref{fig:histogram_M} shows that these two distinct peaks are related to a difference in X-ray properties of early and late M dwarfs.  The median of the $\log(L_\mathrm{X}/L_\mathrm{bol})$ values in each spectral subtype bin rises from values of around $-5$ for M stars earlier than M4 to values above $-4$  for later M stars.

The comparison of the Kaplan-Meier estimator for the  $L_\mathrm{X}/L_\mathrm{bol}$ distributions of M stars (including upper limits) in different spectral subclass bins in the lower part of Fig.~\ref{fig:histogram_M} shows that M stars later than M4 display a distribution toward higher $L_\mathrm{X}/L_\mathrm{bol}$ values, and the bump at high activity levels becomes even more obvious. For earlier stars, the distributions become more and more smooth, and the bump completely vanishes for SpT $<$\;M4.

It is interesting to note that the transition occurs at a SpT of about M3--M4.
 The observed change in activity levels may thus be related to the change in stellar structure and the corresponding dynamo action across the transition from early M-stars with a convective envelope and radiative interior, to fully convective late M-stars, which is expected
to occur around a stellar mass of $M \simeq 0.35\;M_\odot$, corresponding to a SpT of $\simeq$\;M3--M4 \citep[see, e.g.,][and references therein]{2017LRSP...14....4B}.
Stars with SpTs earlier than $\simeq$\;M3 are partly convective and are expected to operate a solar-type $\alpha \Omega$-dynamo, in which the magnetic field is generated in the tachocline at the interface between the radiative core and convective shell \citep[see, e.g.,][for reviews of the solar dynamo]{ solar_dynamo2, solar_dynamo, 2017LRSP...14....4B}.
Stars of SpT M4 and later ($M \la 0.35\;M_\sun$) are fully convective. Since they have no tachocline, they must operate with a different type of dynamo; for example,~an $\alpha^2$-dynamo for which only helical turbulences are the main driver of magnetic activity \citep{alpha2-dynamo, alpha2, 2021A&A...651A..66K, PMS_X_ray2}.

Several previous studies of the X-ray activity of M-type stars have searched for systematic differences between the (partly convective) early M stars and the
(fully convective) later M stars, with mixed results:
 \cite{2015RSPTA.37340259T} concluded that M-type field stars show no obvious change in X-ray activity levels at the
transition to a fully convective structure, whereas
\citet{10pc_XUV} reported higher X-ray activity levels in a sample of nearby fully convective M-type stars.

However, the observed higher X-ray activity levels for fully convective stars in our sample do not allow us to draw immediate strong conclusions about stellar dynamos, since it is 
 well established that the X-ray activity of late-type stars depends on the rotation period, probably via the Rossby number (i.e.,~the ratio between rotation period and the convective turnover time) \citep{activity_relation3, Wright_2011, activity_relation2,activity_relation}. In general, faster rotation leads to higher activity levels, but for periods below about 1~day (or Rossby numbers below about 0.1), a saturation effect limits the X-ray activity to typical levels on the order of $L_\mathrm{X}/L_\mathrm{bol} \sim 10^{-3}$.

As stars spin down with time, their (X-ray) activity levels decrease with time, but the temporal evolution of the spindown process depends strongly on stellar mass \citep[e.g.,][]{2023A&A...675A.180G}. Studies of the rotation periods in large samples of field stars show that the early M-type field stars show a roughly homogeneous distribution of periods (between $\la 1$~day and up to $\sim 100$~days), whereas the (fully convective) late M field stars seem to show a bimodal distribution: many slow rotators with periods of $\ga 100$~days and many fast rotators with periods of $\la 1$~day, but not much in between \citep{rotation_period_mass, rotation_period_mass2, Newton_2016, 2018AJ....156..217N}. 
Therefore, the difference in the X-ray activity levels between early and late M-type stars in our sample
may be reflecting such differences in the rotation rate.

Since measured rotation periods are available for fewer than half of the M dwarfs in our sample,
our current data do not allow us to distinguish between the effects of a change in stellar structure (and the corresponding nature of the dynamo) and possible systematic differences in the rotation periods.
Therefore, this topic needs to be addressed separately, when more complete rotation data become available.

\subsection{Comparison between M-type and G/K-type stars}

\begin{figure}[htbp]
    \centering
    \includegraphics[width=7.5cm]{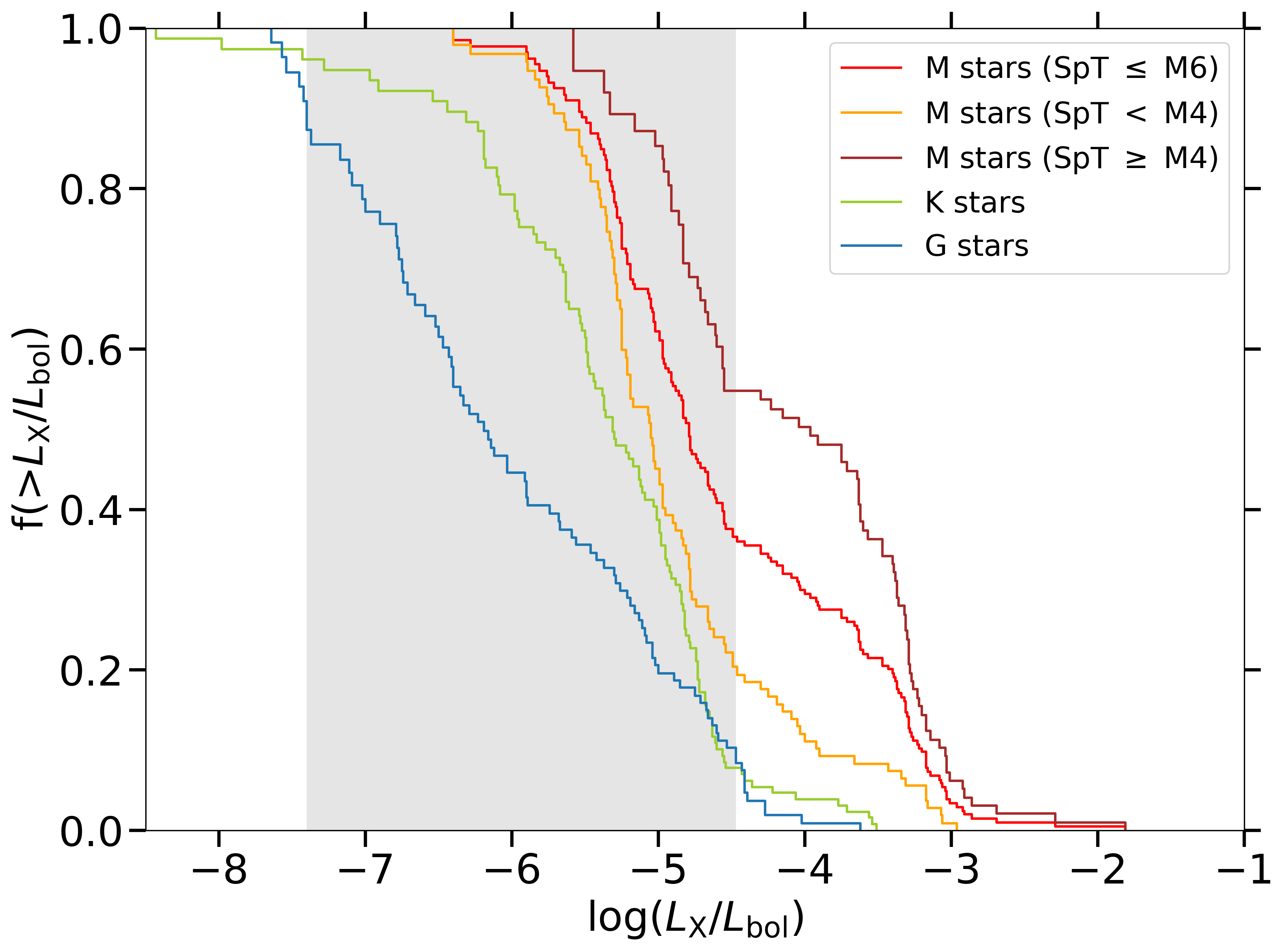}
    \caption{Comparison of the Kaplan-Meier estimators for the $\log(L_\mathrm{X}/L_\mathrm{bol})$ distributions of G, K, and M stars in this work. Distributions for late and early M stars are also plotted. The central 80\% quantile of G stars is shaded in gray.}
    \label{fig:cumulative_comparison_all}
\end{figure}

Fig.~\ref{fig:cumulative_comparison_all} shows a comparison of the distributions of $\log(L_\mathrm{X}/L_\mathrm{bol})$ of this work between different SpTs. 
As was expected, we see that the
M-type stars show systematically higher values of X-ray activity than G-type (and also K-type) stars.
The median values are $\log(L_\mathrm{X}/L_\mathrm{bol})=-4.80$  for the M-type stars, $-5.32$ for the K-type ones, and $-6.20$ for the G-type ones.
It is important to note, however, that the distributions for the three spectral classes overlap strongly.
In order to allow for a meaningful comparison of the different SpTs, we calculated the central 80\% quantile of the $\log(L_\mathrm{X}/L_\mathrm{bol})$ distribution for G stars and considered it to be a ``typical activity range'' for solar-like stars; this quantile ranges
from  $\log(L_\mathrm{X}/L_\mathrm{bol}) = -7.40$
to $\log(L_\mathrm{X}/L_\mathrm{bol}) = -4.47$.

In our sample, 63\% of all M stars have $L_\mathrm{X}/L_\mathrm{bol}$ values within the 80\% quantile of the G stars; only 37\% of all M stars  show higher activity levels.
This means that the vast majority of nearby M stars (no later than M6) show X-ray activity levels that do not exceed those  of the typical activity range for solar-like stars. If we further distinguish between early M stars with SpT $<$\;M4 and late M stars with SpT $\geq$\;M4, then 80\% of the former and 45\% of the latter are within the 80\% quantile of the G stars.

\section{Implications for planet habitability}\label{sec:habitability}

Although M-type stars show systematically higher levels of X-ray activity that solar-like stars, we find that $\ga 60$\% of all nearby M stars [$\leq $\;M6], and 80\% of the early M stars [$<$\;M4] actually have $L_\mathrm{X}/L_\mathrm{bol}$ values that are still within the typical range for G-type (i.e.,~solar-like) stars. This implies that the X-ray flux received by planets around the (vast) majority of (early) M stars is comparable to that commonly received by planets orbiting typical (solar-like) G-type stars.
For planets around these M stars, there is thus no strong reason to assume that the X-ray irradiation would pose significantly more severe problems for their habitability than it would for the habitability of planets orbiting G-type stars.

Another way to quantify  the level of X-ray irradiation for planets around M-type stars is to directly compare the received X-ray flux to that received by Earth from the Sun.
The fractional X-ray luminosity of our Sun (during the maximum of the solar cycle) is  $L_\mathrm{X}/L_\mathrm{bol} \simeq -5.68$ \citep[see][]{Sun_Lx}.
Although most M stars show higher values than this, a large fraction of the M stars exceed this value by just a factor of a few. Considering the distribution in our nearby M star sample, we find that 
only 45\% of all nearby M stars no later than M6, and only 28\% of early M stars with SpT $<$ M4, exceed the solar max $L_\mathrm{X}/L_\mathrm{bol}$ by a factor of more than ten. This implies that most planets around M-type stars are exposed to only moderately enhanced levels of X-ray irradiation, which should have very limited effects on their atmospheres and the habitability.
Considering a limit of 100 times the solar max value, we find that
only 26\% of M stars no later than M6 (and only 10\% of early M stars) exceed this limit.

It is, however, important to keep in mind that the X-ray flux received by a planet is only one piece of the big mosaic of factors determining the habitability of a planet. X-ray and UV irradiation causes ionization, photodissociation, and heating processes in the atmosphere, and can be an important driver of atmospheric escape and loss of planet atmosphere \citep[see, e.g.,][]{Doyon_overview}. Whether planets around active M stars can retain their atmosphere or not also depends strongly on other parameters than $L_\mathrm{X}/L_\mathrm{bol}$, such as planet mass and radius, the chemical composition of the planet atmosphere, the planetary magnetic field, and replenishment through internal (volcanic and tectonic process on the planet) or external (e.g., comet impacts) sources, and whether the planet is tidally locked \citep{Fossati2017, Airapetian_review, planet_interior_heating, planet_interior_heating2}.
Due to the complexity of the problem, atmospheric mass loss of planets around M dwarfs needs to be studied on a
case-by-case basis \citep[see e.g.][]{Foster_2022}. Dedicated studies of the XUV spectra of M-type exoplanet host stars such as the MUSCLES \citep{MUSCLES} and MEGA-MUSCLES \citep{MEGA_MUSCLES} projects, and future studies of rocky planet atmosphere around M dwarfs with JWST \citep{JWST_DDT, Doyon_overview}, will soon shed more light on this important topic.

\section{Conclusion and summary}\label{sec:conclusion}

We have used \textit{Gaia} data to assemble unbiased, volume-limited samples of 205 M stars with SpT $\le$ M6 within 10~pc of the Sun, 129 K stars within 16~pc, and 107 G stars within 20~pc, and collected X-ray data from \textit{Chandra}, \textit{XMM-Newton}, eROSITA, and ROSAT for these stars to construct the distribution functions for the absolute and relative X-ray luminosities ($L_\mathrm{X}/L_\mathrm{bol}$) of these stars. The X-ray detection completeness of our samples is 85\% for M stars, 86\% for K stars, and 80\% for G stars.

The X-ray activity $\log(L_\mathrm{X}/L_\mathrm{bol})$ of M stars in our sample shows a bimodal distribution: there is one peak at around $-5$, and a second peak between $-4$ and $-3$ that is largely caused by late (SpT $\geq$ M4) M dwarfs.
The systematically higher X-ray activity of late M dwarfs compared to early ones in our sample may be related to the transition from a partly convective structure to a fully convective one and the corresponding change in stellar dynamo that is expected around SpT M3--4, but it may also be caused by possible systematic differences in the rotation periods between early and late M stars.

Comparing the distributions of $L_\mathrm{X}/L_\mathrm{bol}$ of G-, K-, and M-type stars shows that, although $L_\mathrm{X}/L_\mathrm{bol}$ systematically increases toward later SpTs, 63\% of all M stars in our sample have $L_\mathrm{X}/L_\mathrm{bol}$ values within the central 80\% quantile for G stars.
If we further distinguish between early M stars with SpT $<$ M4 and late M stars with SpT $\geq$ M4, then these fractions become 80\% for the former and 45\% for the latter.

Although some (particularly active) M-type stars show X-ray activity levels that would probably be problematic for potential life forms on habitable-zone planets (or that would have destroyed a planetary atmosphere very quickly), the clear majority of stars with SpTs $<$\,M4 
as well as a large fraction of the later M-type stars (M4--M6)
produce only moderately enhanced X-ray irradiation levels, and thus
should not be excluded as potential hosts of habitable planets.

\begin{acknowledgements}
The work of T.P.~is partly supported by  the Excellence Cluster ORIGINS which is funded by the Deutsche Forschungsgemeinschaft (DFG, German Research Foundation) under Germany’s Excellence Strategy - EXC-2094 - 390783311.
This research has made use of the SIMBAD database, operated at CDS, Strasbourg, France. The scientific results reported in this article are based in part on data obtained from the Chandra Data Archive and observations made by the Chandra X-ray Observatory. This research has made use of software provided by the Chandra X-ray Center (CXC) in the application packages CIAO. This research has also made use of data obtained from the 4XMM XMM-Newton serendipitous source catalog compiled by the XMM-Newton Survey Science Centre, and archival data of the ROSAT space mission. This work uses  data from eROSITA, the soft X-ray instrument aboard SRG, a joint Russian-German science mission supported by the Russian Space Agency (Roskosmos), in the interests of the Russian Academy of Sciences represented by its Space Research Institute (IKI), and the Deutsches Zentrum für Luft- und Raumfahrt (DLR). The SRG spacecraft was built by Lavochkin Association (NPOL) and its subcontractors, and is operated by NPOL with support from the Max Planck Institute for Extraterrestrial Physics (MPE). The development and construction of the eROSITA X-ray instrument was led by MPE, with contributions from the Dr. Karl Remeis Observatory Bamberg \& ECAP (FAU Erlangen-Nuernberg), the University of Hamburg Observatory, the Leibniz Institute for Astrophysics Potsdam (AIP), and the Institute for Astronomy and Astrophysics of the University of Tübingen, with the support of DLR and the Max Planck Society. The Argelander Institute for Astronomy of the University of Bonn and the Ludwig Maximilians Universität Munich also participated in the science preparation for eROSITA. The eROSITA data shown here were also partly processed using the eSASS software system developed by the German eROSITA consortium. This research has made use of data and/or software provided by the High Energy Astrophysics Science Archive Research Center (HEASARC), which is a service of the Astrophysics Science Division at NASA/GSFC.
\end{acknowledgements}

\bibliographystyle{aa}
\bibliography{reference}

\begin{appendix}

\onecolumn

\section{References used in the catalog}\label{appendix:reference}

\begin{table*}[h!]
 \caption{\textit{Chandra} archival observations from which we determined $L_\mathrm{X}$ with CIAO.}
 
 \centering
 \begin{tabular}{|l|l|l|l|l|l|l|}
 \hline
 Star & SpT & distance [pc] & Instrument & Grating & Exposure time [ks] & Observation ID \\
 \hline
 \hline
 GJ~65A/B & M5/M6 & 2.7 & HRC-S & LETG & 56.6 & ~\dataset[22876]{https://doi.org/10.25574/22876}\\
 GJ~54.1 & M4 & 3.7 & ACIS-S & NONE & 9.93 & ~\dataset[22291]{https://doi.org/10.25574/22291}\\
 GJ~628 & M3 & 4.3 & ACIS-S & NONE & 38.51 & ~\dataset[20163]{https://doi.org/10.25574/20163}\\
 GJ~166C & M4.5 & 5.0 & ACIS-S & NONE & 5 & ~\dataset[13644]{https://doi.org/10.25574/13644}\\
 GJ~169.1A & M4 & 5.5 & ACIS-S & NONE & 4.98 & ~\dataset[13643]{https://doi.org/10.25574/13643}\\
 GJ~644AB/\newline GJ~643E & M3/\newline M3.5 & 6.2 & ACIS-S & NONE & 9.04 & ~\dataset[615]{https://doi.org/10.25574/615}\\
 GJ~1156 & M4.5 & 6.5 & ACIS-S & NONE & 19.81 & ~\dataset[27611]{https://doi.org/10.25574/27611}\\
 GJ~569A & M1 & 9.9 & ACIS-S & NONE & 24.86 & ~\dataset[4470]{https://doi.org/10.25574/4470}\\
 \hline
 GJ~570A/\newline BC & K4/\newline M1 & 5.9 & ACIS-S & NONE & 39.63 & ~\dataset[8904]{https://doi.org/10.25574/8904}\\
 GJ~663A/B & K1/K2 & 6.0 & HRC-S & LETG & 77.37 & ~\dataset[4483]{https://doi.org/10.25574/4483}\\
 GJ~667A & K3 & 6.8 & ACIS-S & NONE & 18.2 & ~\dataset[17317]{https://doi.org/10.25574/17317}\\
 GJ~451 & K1 & 9.3 & ACIS-S & NONE & 32.79 & ~\dataset[9931]{https://doi.org/10.25574/9931}\\
 GJ~615 & K3 & 13.6 & ACIS-S & NONE & 32.5 & ~\dataset[7441]{https://doi.org/10.25574/7441}\\
 \hline
 GJ~19 & G2IV & 7.5 & ACIS-S & NONE & 4.91 & ~\dataset[12337]{https://doi.org/10.25574/12337}\\
 GJ~759 & G5 & 14.9 & HRC-I & NONE & 8.99 & ~\dataset[22309]{https://doi.org/10.25574/22309}\\

 \hline

 \end{tabular}

 \label{tab:chandra_observation}
\end{table*}

\noindent \textbf{\textit{SpT}}. \cite{2017ApJ...840...87R, 1974ApJS...28....1J, 1977IBVS.1323....1H, 1984AJ.....89.1229R, 1984ApJS...55..657C, 1985AJ.....90..817B, 1985ApJS...59..197B, 1986AJ.....92..139S, 1986AJ.....92.1424T, 1990ApJ...360..490R, 1995A&AS..110..367N, 1995AJ....110.1838R, 1996AJ....112.2799H, 1997AJ....113.2246R, 1997PASP..109..849G, 2001A&A...368..569V, 2001ApJ...560..390L, 2002AJ....123.2002H, 2002AJ....124..519R, 2002ApJ...564..466G, 2002ApJ...572L..79L, 2003AJ....125.1598L, 2003AJ....126.2048G, 2003AJ....126.2421C, 2003AJ....126.3007R, 2004A&A...415..265H, 2004AJ....128..463R, 2005A&A...440.1061L, 2005A&A...442..211S, 2005ApJ...634.1336B, 2006A&A...460..695T, 2006A&A...460L..19M, 2006AJ....132..161G, 2006AJ....132..866R, 2006AJ....132.2360H, 2006MNRAS.366L..40P, 2007AJ....133..439C, 2007AJ....133.2825R, 2007ApJ...654..558D, 2007PhDT.........2B, 2008A&A...484..575J, 2008MNRAS.384..150L, 2009ApJ...699..649S, 2009ApJ...704..975J, 2011AJ....141..117J, 2011ApJS..197...19K, 2012A&A...540A.131V, 2012A&A...545A..85R, 2012AJ....144...99D, 2012ApJ...748...93R, 2012RAA....12..443B, 2013A&A...556A..15R, 2013AJ....145..102L, 2013AJ....146..161M, 2014A&A...568A..54R, 2014AJ....147...20N, 2014AJ....147...21J, 2014AJ....147...26D, 2014AJ....147..160M, 2014AJ....148...91L, 2014MNRAS.443.2561G, 2015A&A...577A.128A, 2015AJ....149..104B, 2015AJ....149..106D, 2015ApJ...800...95L, 2015ApJ...804...96G, 2015ApJ...806...62B, 2015ApJ...812....3W, 2015RAA....15.1095L, 2016AJ....151..169R, 2016ApJS..224...36K, 2017A&A...600A..19P, 2017ApJS..231...15D, 2018A&A...613A..26S, 2018AJ....155..265H, 2019A&A...625A..68S, 2019AJ....157...63K, 2019AJ....158...75H, 2019ApJ...876..115S, 2019ApJ...877...60B, 2019ApJ...883..205B, 2019MNRAS.485.2167G, 2019NatAs...3.1099G, 2020A&A...637A..43K, 2020ApJ...892...31B, 2021AJ....161..172D, 1961RGOB...48..389E, 1962ApJ...136..793W, 1967AJ.....72.1334C, 1976AJ.....81..245E, 1989ApJS...71..245K, 1993yCat.3135....0C, 1995AJ....109..332M, 1999MSS...C05....0H, 2018ApJS..238...29P, 1935ApJ....81..187A, 1957MNRAS.117..534E, 1981ApJS...45..437A}\\

\noindent \textbf{\textit{$L_\mathrm{X}$ from pointed observations}}\footnote{\cite{2018MNRAS.479.2351W} contains not only measurements/upper limits from \textit{Chandra} but also from ROSAT. For 4 M stars ROSAT $L_\mathrm{X}$ are collected from this paper.}. \cite{2011A&A...534A.133F, 2023AJ....165..195B, 2018MNRAS.479.2351W, 2017MNRAS.471.1012B, 2008ApJ...676.1307B, 2022AJ....164..206B, 2022MNRAS.512.1751P, 2014ApJ...785....9W, 2022ApJS..263...41A, 2022ApJ...933L..17M, 2012IAUS..286..335S, 2008ApJ...687.1339K, 2021ApJ...921..122M, 2012ApJ...753...76W, 2023AJ....166..167M, 2022MNRAS.513.4380I, 2009A&A...508.1417P, GJ486}

\FloatBarrier 
\twocolumn


\section{\textit{Gaia} bolometric correction}\label{app: BC}

\begin{figure}[htbp]
    \centering
    \includegraphics[width=\hsize]{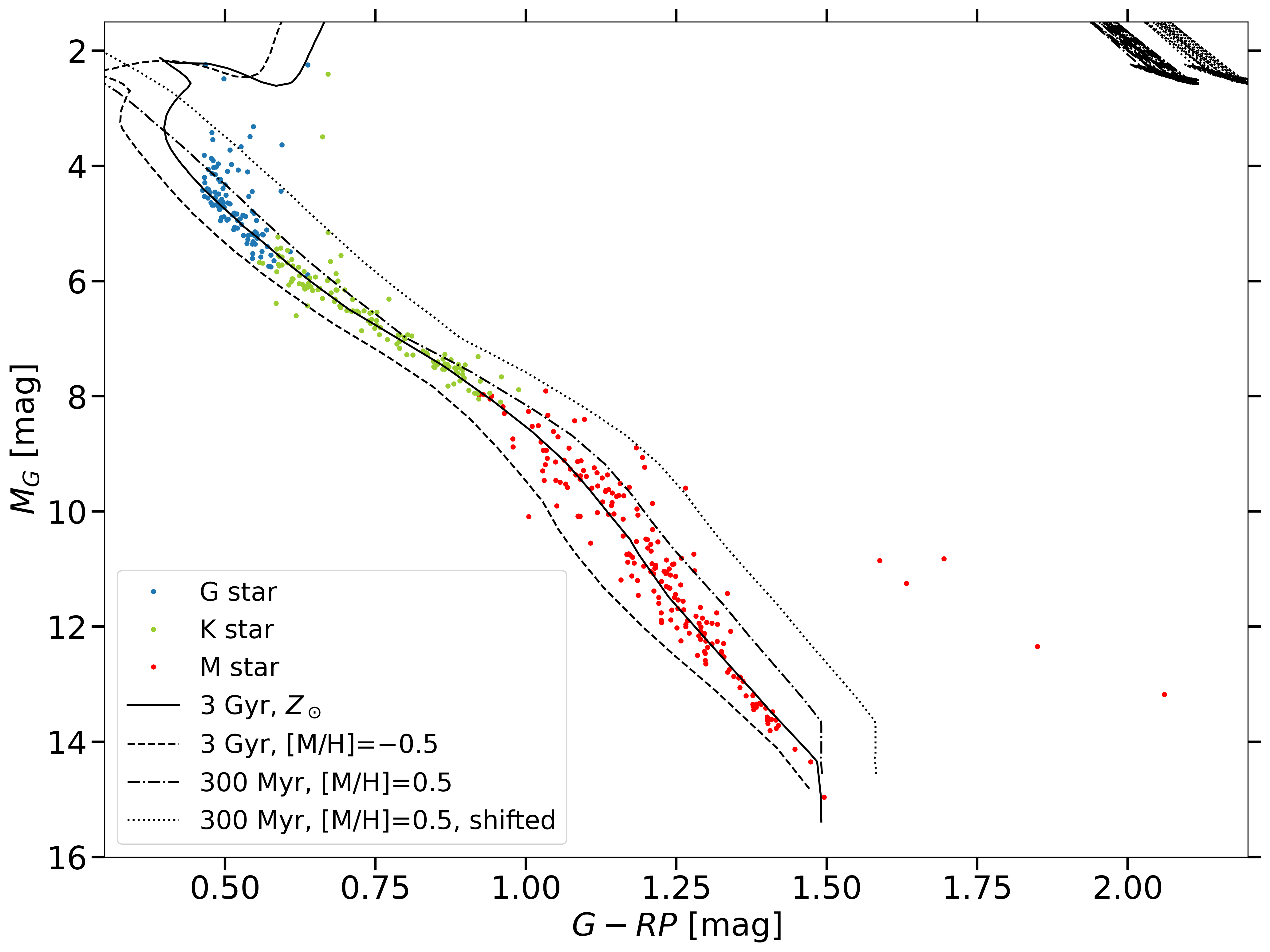}
    \caption{Color-magnitude diagram of the GKM stars in the samples of this work, overplotted with isochrones of different ages and metallicities.}
    \label{fig:CMD_fin}
\end{figure}

For stars that do not have a documented $L_\mathrm{bol}$ in the \textit{Gaia} astrophysical parameter catalog, we used \textit{Gaia} BC routine to calculate $BC_G$ based on the stellar effective temperature $T_\text{eff}$, gravitational acceleration $\log g$, metallicity [Fe/H] and alpha-enhancement [$\alpha$/Fe]  \citep{GAIADR3_2}. $L_\mathrm{bol}$ of the stars were then calculated based on the following equation:
\begin{equation}
    L_\mathrm{bol}/L_\sun = 10^{-(BC_G + M_G-4.74)/2.5}
,\end{equation}
where the solar bolometric magnitude $M_{\text{bol,}\sun}=4.74\,\mathrm{mag}$ is adopted \citep{Mbol_solar}.

As [$\alpha$/Fe] is only estimated for a few stars, it is assumed as 0 for all stars in \textit{Gaia} DR3 and we adopted this setting. $\log g = 5.0$ was assumed for M stars, since typical log $g$ for M dwarf is between 4.5 and 5.5 \citep{logg}. For K and G stars, the median value of the measured $\log g$ in the fourth update of the Geneva Copenhagen Survey \citep{GCS, GCS_advance} which contains stellar parameter measurements of over 16000 FGK dwarfs was used as an approximation: for G stars $\log g = 4.46$ and for K stars $\log g = 4.59$ was assumed. Finally, a five-order polynomial function of $T_\text{eff}$ dependent on $M_G$ was fitted to the values of the Mamejek table and used as approximation for the stellar temperatures in our sample:
\begin{equation}
    \begin{split}
    T_\text{eff} = 0.105 \times M_G^5 -5.622 \times M_G^4 + 111.974 \times M_G^3 \\
    - 1001.961 \times M_G^2 
    + 3502.718 \times M_G + 2277.291
    \end{split}
.\end{equation}

However, for stars that have peculiar \textit{Gaia} colors (for instance due to a crowded field or binary star contamination) $L_\mathrm{bol}$ calculated from the \textit{Gaia} BC routine might be poorly determined. The color-magnitude diagram (CMD) of all stars in the samples of this work (excluding the X-ray non-resolved secondary components) overplotted with isochrones are shown in Fig.~\ref{fig:CMD_fin}.  We used PARSEC v1.2s stellar models \citep{parsec} to generate an isochrone for an age of 3~Gyr, solar metallicity, no reddening, and a minimal mass of 0.09~$M_\sun$, which is plotted as the solid line in Fig.~\ref{fig:CMD_fin}. An isochrone with 3~Gyr and metallicity of $\mathrm{[M/H]}=-0.5$ fits the left edge of the MS very well (the dashed line in Fig.~\ref{fig:CMD_fin}), and an isochrone with 300~Myr and $\mathrm{[M/H]}=0.5$ fits for M stars with $M_G < 12\,\mathrm{mag}$ and K, G stars the right edge of the MS very well (the dash-dotted line in Fig.~\ref{fig:CMD_fin}). 300 Myr and $\mathrm{[M/H]}=0.5$ were chosen because it is the maximal metallicity that PARSEC v1.2s can model, and 300~Myr is approximately the time that a 0.1~$M_\sun$ star reaches the MS, as we expect only young stars to have such a high metallicity.

All stars left to the isochrone with $\mathrm{[M/H]}=-0.5$ on the CMD are considered too blue and have peculiar \textit{Gaia} colors. Some M stars with $M_G > 12\,\mathrm{mag}$ lie rightward to the isochrone with 300~Myr and $\mathrm{[M/H]}=0.5$ but can still be considered to be belong to the MS and have valid \textit{Gaia} $L_\mathrm{bol}$. To include these stars in the analysis this isochrone was shifted empirically to the right for 0.09~mag (the dotted line in Fig.~\ref{fig:CMD_fin}), and all stars right to this shifted isochrone are considered too red and have peculiar \textit{Gaia} colors. These stars are flagged in our samples (see Table \ref{tab:sample} for column description of our sample), and their $L_\mathrm{bol}$ were adopted from the Mamajek table based on their SpT. If there are no SpT measurements from the literature or if they are subgiants, then their $M_G$ were used as references to look up in the table.


\onecolumn

\section{NEXXUS2 matches with large separation}\label{appendix:nexxus}

Three entries in NEXXUS2  have a separation larger than $40^{\arcsec}$ between the stellar position and the X-ray source position:

For GJ~1103, a M4.5 dwarf at 9.3~pc, the X-ray source and the stellar position has a large offset of $115^{\arcsec}$, and  for GJ~785, a K2 star at 8.8~pc, the X-ray source and stellar position have an offset of $62^{\arcsec}$. Despite of the large separations, these two nearby stars are more likely to be the counterparts of these X-ray sources, as there are no other bright optical/infrared sources within the field. 

Considering the X-ray hardness ratios (HR), fhe former X-ray source has HR1 of $1\pm 0.61$ and HR2 of $-0.46 \pm 0.54$ and the latter X-ray source has HR1 of $-1\pm 0.37$ and HR2 of $-1 \pm 0$ according to NEXXUS2. These values are quite typical hardness ratios for stars which emit relatively soft X-rays. In the alternative hypothesis, that the X-ray sources are not the nearby stars but optically very faint quasars, one would expect substantially harder X-ray hardness ratios.
Therefore, we consider these two matches in NEXXUS2 to be true matches.

Another X-ray source is detected between GJ~653 (K5 star at 10.5~pc) and its companion GJ~654 (M1.5). It has HR1 of $-0.33\pm 0.4$ and HR2 of $-0.22 \pm 1.04$. With the same reasoning as in previous cases, it is more likely that the X-ray emission is emitted by one of the two nearby stars. This X-ray source was thus assigned to GJ~654 due to the smaller separation.

\begin{figure*}[h!]
    \centering
    \includegraphics[width=0.31\linewidth]{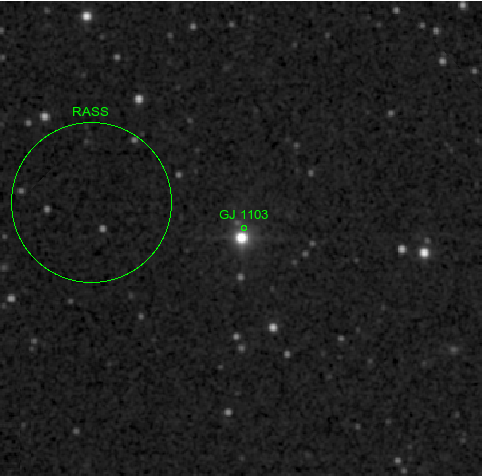}
    \includegraphics[width=0.31\linewidth]{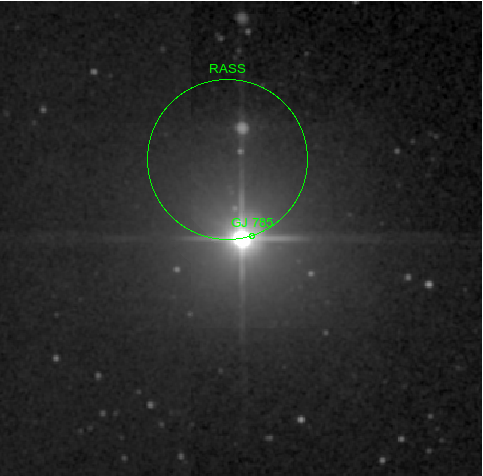}
    \includegraphics[width=0.31\linewidth]{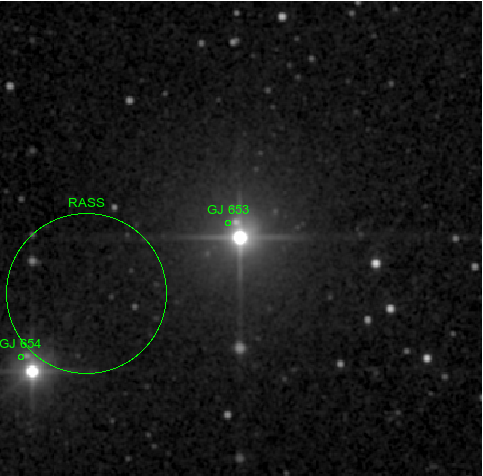}
    \caption{2MASS \textit{J}-band images of the three NEXXUS2 matches with separation larger than $40^{\arcsec}$: GJ~1103 (left), GJ~785 (middle) and GJ~653/GJ~654 (right). The large circle has $60^{\arcsec}$ radius with center being the coordinate of the X-ray source. Small circles have $2^{\arcsec}$ radius with center being the \textit{Gaia} coordinates of stars extrapolated back to epoch 1990.}
    \label{fig:NEXXUS_match}
\end{figure*}

\FloatBarrier 
\twocolumn

\onecolumn
\section{Upper limits from RASS}\label{app: upper_limit}

\begin{figure*}[h!]
    \centering
    \includegraphics[width=0.49\linewidth]{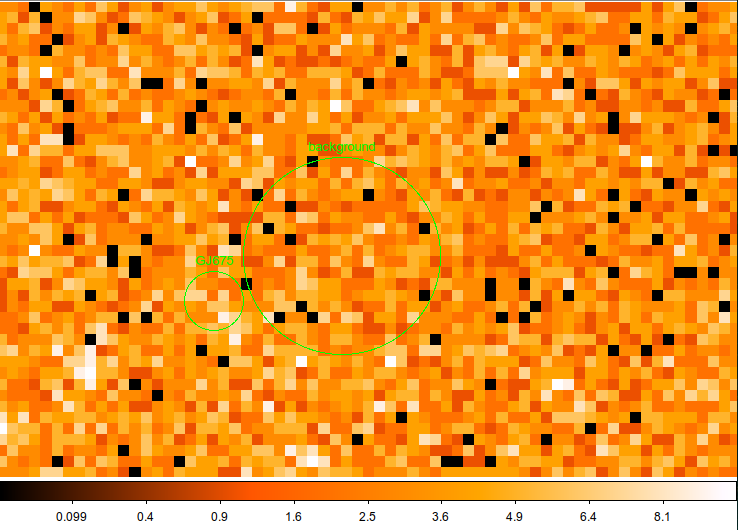}
    \includegraphics[width=0.49\linewidth]{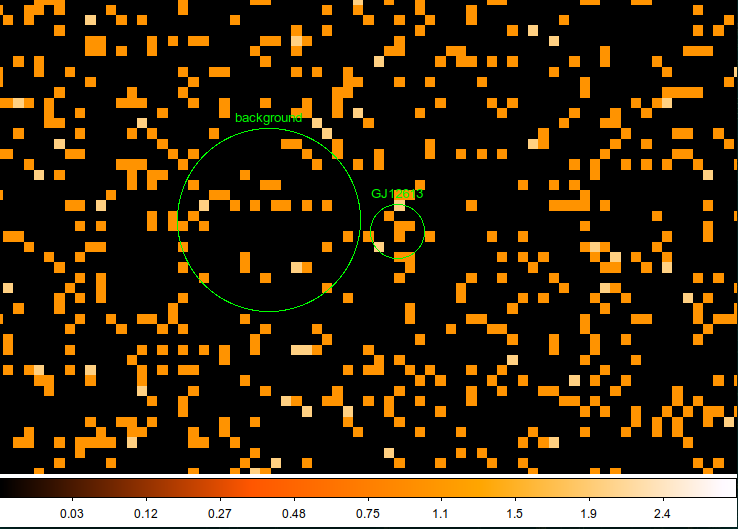}
    \caption{ROSAT RASS images of the stars for which we calculated their flux upper limits: GJ~675 (left) and GJ~12613 (PM~J18057-1422, right). The source region with $120^{\arcsec}$ radius and the background region with $400^{\arcsec}$ radius are shown.}
    \label{fig:upperlimit}
\end{figure*}

There are two stars for which we calculated their flux upper limits from ROSAT RASS images: GJ~675 (K0 star at 12.8~pc) 
and GJ~12613 (or PM~J18057-1422 in SIMBAD, M5.5 star at 9.5~pc), as shown in Fig.~\ref{fig:upperlimit}. After visual inspection of the RASS images, the source region and the background region were defined as circles with radius of $120^{\arcsec}$ and $400^{\arcsec}$. The center of the source region is the stellar position extrapolated back to the epoch of 1990 with \textit{Gaia} proper motions, and the center of the background region was chosen close to the source region. The photon counts within these two regions and the background value of the source region are listed in Table \ref{tab:upper_limit}. With these measurements the upper limit photon count with 90\% confidence level was calculated and converted to X-ray flux with the conversion factor defined by NEXXUS in the end.

\begin{table*}[h!]
 \caption{Properties of stars for which we calculated their flux upper limits from RASS images. See text for details about the columns.}
 
 \centering
 \begin{tabular}{|p{2cm}|p{2cm}|p{2cm}|p{2cm}|p{2cm}|p{2cm}|p{2cm}|}
 \hline
 Star & total count in background region & scaled background count  &count in source region  & upper limit photon count & exposure time [s] & $L_\mathrm{X}$ upper \newline limit [$\mathrm{erg\,s}^{-1}$]\\
 \hline
 \hline
 GJ~675 & 812  & 73.08 & 96  & 39.76 & 5311.15 & $8.79 \times 10^{26}$ \\
 \hline
 GJ~12613 & 32  & 2.88 & 10  & 13.31 & 273.803 & $3.13 \times 10^{27}$\\
 
 \hline

 \end{tabular}
 
 \label{tab:upper_limit}
\end{table*}

\FloatBarrier 
\twocolumn

\onecolumn

\section{Coronal temperatures}\label{appendix:coronal_T}

Tables \ref{tab:kT} and \ref{tab:kT2} list coronal temperatures we adopted directly from literature. For some stars, their spectra were fitted with multitemperature models with emission measures (EM) describing the contribution from each temperature component. In these cases, we calculated the EM-weighted temperature ($kT_\text{wgt}$) as their coronal temperatures (see Table \ref{tab:kT3}). The spectra of GJ~551C, GJ~876 and GJ~799A were separated into quiescence/low and flare/high components and fitted separately in the literature, the final coronal temperatures we used are the average of the $kT_\text{wgt}$ of the two components.

\begin{table*}[h!]
 \caption{M stars for which we adopted coronal temperatures from literature directly.}
 
 \centering
 \begin{tabular}{|p{1.5cm}|p{1cm}|p{2cm}|p{2cm}|p{4cm}|}
 \hline
 star & SpT & distance [pc] & $kT$ [keV] & reference \\
 \hline
 \hline
 GJ~699 & M3.5 & 1.8 & 0.54 & {\cite{2023AJ....165..195B}}\\
 GJ~65A & M5 & 2.7 & 0.34 & {\cite{2003ApJ...589..983A}}\\
 GJ~65B & M6 & 2.7 & 0.34 & {\cite{2003ApJ...589..983A}}\\
 GJ~15A & M1 & 3.6 & 0.45 & {\cite{2023AJ....165..195B}}\\
 GJ~15B & M3.5 & 3.6 & 0.56 & {\cite{2023AJ....165..195B}}\\
 GJ~1245A & M5.5 & 4.7 & 0.3 & {\cite{2014ApJ...785....9W}}\\
 GJ~832 & M1.5 & 5.0 & 0.23 & {\cite{2023AJ....165..195B}}\\
 GJ~873 & M4 & 5.1 & 0.612 & {\cite{2022MNRAS.512.1751P}}\\
 GJ~3323 & M4 & 5.4 & 0.76 & {\cite{2018MNRAS.479.2351W}}\\
 GJ~754 & M4.5 & 5.9 & 0.59 & {\cite{2018MNRAS.479.2351W}}\\
 GJ~285 & M4 & 6.0 & 0.6463 & {\cite{2022MNRAS.512.1751P}}\\
 GJ~581 & M3 & 6.3 & 0.26 & {\cite{2023AJ....165..195B}}\\
 GJ~3193A & M3 & 6.9 & 0.63 & {\cite{2022AJ....164..206B}}\\
 GJ~3192BC & M2.5 & 6.9 & 0.77 & {\cite{2022AJ....164..206B}}\\
 GJ~1286 & M5.5 & 7.2 & 0.52 & {\cite{2018MNRAS.479.2351W}}\\
 GJ~667C & M1.5 & 7.2 & 0.41 & {\cite{2023AJ....165..195B}}\\
 GJ~486 & M3.5 & 8.1 & 0.19 & {\cite{GJ486}} \\
 GJ~849 & M3.5 & 8.8 & 0.32 & {\cite{2023AJ....165..195B}}\\
 GJ~867A & M2 & 8.9 & 0.5515 & {\cite{2022MNRAS.512.1751P}}\\
 GJ~367 & M1 & 9.4 & 0.16 & {\cite{GJ367}}\\
 GJ~176 & M2 & 9.5 & 0.31 & {\cite{2023AJ....165..195B}}\\
 GJ~1256 & M4.5 & 9.5 & 0.69 & {\cite{2018MNRAS.479.2351W}}\\
 GJ~3253 & M5 & 9.7 & 0.3 & {\cite{2018MNRAS.479.2351W}}\\
 GJ~803 & M0.5 & 9.7 & 0.6463 & {\cite{2022MNRAS.512.1751P}}\\
 GJ~436 & M2.5 & 9.8 & 0.39 & {\cite{2023AJ....165..195B}}\\
 
 \hline

 \end{tabular}
 
 \label{tab:kT}
\end{table*}

\begin{table*}[h!]
 \caption{G/K stars for which we adopted coronal temperatures from literature directly.}
 
 \centering
 \begin{tabular}{|p{1.5cm}|p{1cm}|p{2cm}|p{2cm}|p{4cm}|}
 \hline
  star & SpT & distance [pc] & $kT$ [keV] & reference \\
 \hline
 \hline
 GJ~166A & K0.5 & 5.0 & 0.19 & {\cite{2017MNRAS.471.1012B}}\\
 GJ~117 & K1.5 & 10.4 & 0.51 & {\cite{2022MNRAS.512.1751P}}\\
 GJ~370 & K6 & 11.3 & 0.25 & {\cite{2023AJ....165..195B}}\\
 GJ~2046 & K2.5 & 12.9 & 0.15 & {\cite{2023AJ....165..195B}}\\
 GJ~222 & G0 & 8.7 & 0.517 & {\cite{2022MNRAS.512.1751P}}\\
 GJ~137 & G5 & 9.3 & 0.49 & {\cite{2022MNRAS.512.1751P}}\\
 GJ~672 & G0 & 14.6 & 0.05 & {\cite{2012IAUS..286..335S}}\\
 GJ~599A & G7 & 15.3 & 0.44 & {\cite{2017MNRAS.471.1012B}}\\
 GJ~882 & G2 & 15.5 & $<$0.09 & {\cite{2009A&A...508.1417P}}\\
 GJ~777A & G7 & 16.0 & 0.26 & {\cite{2022MNRAS.513.4380I}}\\

 \hline

 \end{tabular}
 
 \label{tab:kT2}
\end{table*}

\begin{table*}[h!]
 \caption{Stars for which we calculated the emission measure-weighted coronal temperatures $kT_\text{wgt}$. 
 Because only the ratio between EM of each component matters for calculating $kT_\text{wgt}$, they are listed as unitless in the table.}
 \begin{tabular}{|p{1.5cm}|p{2cm}|p{2cm}|p{2cm}|p{2cm}|p{2cm}|p{2cm}|p{2cm}|p{2cm}|}
  \cline{1-8}
    &GJ~551C \newline(quiescence)  &GJ~551C \newline (flare) & GJ~729 & GJ~674 &GJ~1245B$^*$ &GJ~876 (low)  &GJ~876\newline (high)\\
    \cline{1-8}
   SpT & M5.5 & &M3.5 &M2 &M5.5& M4 &\\
   distance [pc] & 1.3  &  &3.0 & 4.6 &4.7& 4.7  &\\
   $kT_1$ [keV]& 0.14 & 0.14 &0.15 &0.12 &0.22& 0.14 & 0.24\\
   $\text{EM}_1$ & 0.16& 0.17 &1.83 &1.28 & & 1.3 & 1.8\\
   $kT_2$ [keV]& 0.4& 0.4 &0.34 &0.27 &0.87& 0.8& 0.68\\
    $\text{EM}_2$ & 0.35 & 1.42 &3.32 &1.84 && 0.37 & 6.5\\
   $kT_3$ [keV]& 1.0 & 1.0 &0.75 &0.81 &&&\\
    $\text{EM}_3$ & 0.01 & 1.1  &2.71 &0.65 &&&\\
    \cline{1-8}
   $kT_\text{wgt}$ [keV] & 0.33 & 0.63 &0.44 &0.31 & 0.545& 0.29 & 0.58\\
   reference &  {\cite{2011A&A...534A.133F}}  & {\cite{2011A&A...534A.133F}} &{\cite{2023AJ....165..195B}} &{\cite{2023AJ....165..195B}} & {\cite{2014ApJ...785....9W}}&  {\cite{2023AJ....165..195B}}  & {\cite{2023AJ....165..195B}}\\
  \cline{1-8}
  \hline
  \hline
  &GJ~799A \newline (quiescence)  &GJ~799A \newline (flare) & GJ144 & GJ~820A & GJ~820B & GJ783A\\
    \cline{1-7}
   SpT & M4.5 & & K2 & K5 & K7 & K2.5\\
   distance [pc] & 9.9  &  & 3.2 & 3.5 & 3.5 & 6.0\\
   $kT_1$ [keV]& 0.2723 & 0.2508 & 0.12 & 0.21 & 0.19 & 0.17\\
   $\text{EM}_1$ & 2.34 & 2.71 & 53 & 4.09 & 1.71 & 5.31\\
   $kT_2$ [keV]& 0.6696& 0.7385 & 0.32 & 0.79 & 0.67 & 0.76\\
    $\text{EM}_2$ & 4.85 & 4.27 & 78.4 & 1.75 & 5.06 & 1.45\\
   $kT_3$ [keV]& 1.982 &2.9213 & 0.7  & &&\\
    $\text{EM}_3$ & 4.99 & 12.47   & 22.2 & &&\\
    \cline{1-7}
   $kT_\text{wgt}$ [keV] & 1.13 & 2.07 & 0.31 & 0.3838 & 0.55&0.30\\
   reference &  {\cite{GJ799}}  & {\cite{GJ799}} & {\cite{2023AJ....165..195B}}& {\cite{2017MNRAS.471.1012B}}&  {\cite{2017MNRAS.471.1012B}}&{\cite{2017MNRAS.471.1012B}}\\
   \cline{1-7}
 
 \end{tabular}

 \bigskip
 
 $^*$EM is not listed in the paper. $\text{EM}_1 = \text{EM}_2= 0.5$ is thus assumed for calculation.

 \label{tab:kT3}
\end{table*}

\FloatBarrier 
\twocolumn

\onecolumn

\section{Sample content}\label{appendix:sample}

Table \ref{tab:sample} contains the column description of our samples, with the example of Proxima Centauri. The full tables are available at the CDS.

\begin{table*}[h!]
\caption{Content of the samples of this work, with the example of Proxima Centauri.}
\label{tab:sample}

\begin{tabular}{|p{4cm}|p{1cm}|p{7cm}|p{4cm}|}

\hline
  Column name & Unit & Description & Example \\
 \hline
 \hline
 \texttt{CNS5\_id} & & designation in CNS5 & 3591\\
 \texttt{gj\_id} & & Gliese-Jahrei\ss \; number & 551\\
 \texttt{other\_name} & & name of stars that do not have \texttt{CNS5\_id} and \texttt{gj\_id} & \\
 \texttt{component\_id} & & component in a multiple system & C \\
 \texttt{RA} & deg & right ascension (Gaia DR3)& 217.392321472009\\
 \texttt{DEC} & deg & declination (Gaia DR3)& -62.6760751167667\\
 \texttt{GAIA\_peculiar\_color} & & flag for stars with peculiar \textit{Gaia} colors  & 0 \\
 \texttt{SpT} & & SpT & M5.5\\
 \texttt{r\_SpT} & & reference of \texttt{SpT} & 1995AJ....110.1838R\\
 \texttt{Gaia\_SpT} & & SpT according to $M_G$ which is different from literature SpT & \\
 \texttt{distance} & pc & \textit{Gaia} DR3 distance to the Sun & 1.30\\
 \texttt{Lx\_eROSITA} & $\mathrm{erg\,s}^{-1}$ & X-ray luminosity according to eROSITA & 6.86E+026\\
 \texttt{Lx\_eROSITA\_uerr} & $\mathrm{erg\,s}^{-1}$ & upper error of \texttt{Lx\_eROSITA} & 3.09E+025\\
 \texttt{Lx\_eROSITA\_lerr} & $\mathrm{erg\,s}^{-1}$ & lower error of \texttt{Lx\_eROSITA} & 3.02E+025\\
 \texttt{eROSITA\_upp\_lim} & & flag for \texttt{Lx\_eROSITA} as upper limit value & 0\\
 \texttt{Eband\_Lx\_eROSITA}& keV & energy band of \texttt{Lx\_eROSITA} and its errors & 0.2-2.3\\
 \texttt{r\_Lx\_eROSITA}&&reference of \texttt{Lx\_eROSITA} and its errors & eRASS1\\
 \texttt{Lx\_ROSAT} & $\mathrm{erg\,s}^{-1}$ & X-ray luminosity according to ROSAT & 6.58E+026\\
 \texttt{Lx\_ROSAT\_uerr} & $\mathrm{erg\,s}^{-1}$ & upper error of \texttt{Lx\_ROSAT} & 6.91E+024\\
 \texttt{Lx\_ROSAT\_lerr} & $\mathrm{erg\,s}^{-1}$ & lower error of \texttt{Lx\_ROSAT} & 6.91E+024\\
 \texttt{eROSAT\_upp\_lim} & & flag for \texttt{Lx\_ROSAT} as upper limit value & 0\\
 \texttt{Eband\_Lx\_ROSAT}& keV & energy band of \texttt{Lx\_ROSAT} and its errors & 0.1-2.4\\
 \texttt{instrument\_Lx\_ROSAT} && camera of ROSAT used\tablefootmark{a} & PSPC \\
 \texttt{r\_Lx\_ROSAT}&&reference of \texttt{Lx\_ROSAT} and its errors & NEXXUS\\
 \texttt{Lx\_other} & $\mathrm{erg\,s}^{-1}$ & X-ray luminosity according to other telescopes (\textit{Chandra} or \textit{XMM-Newton}) &6.49E+026 \\
 \texttt{Lx\_other\_uerr} & $\mathrm{erg\,s}^{-1}$ & upper error of \texttt{Lx\_other} & \\
 \texttt{Lx\_other\_lerr} & $\mathrm{erg\,s}^{-1}$ & lower error of \texttt{Lx\_other} & \\
 \texttt{Lx\_other\_upp\_lim} & & flag for \texttt{Lx\_other} as upper limit value & 0\\
 \texttt{Eband\_Lx\_other}& keV & energy band of \texttt{Lx\_other} and its errors & 0.2-10\\
 \texttt{instrument\_Lx\_other} && camera of other telescopes used & EPIC\\
 \texttt{r\_Lx\_other}&&reference of \texttt{Lx\_other} and its errors &2011A\&A...534A.133F\\
 \texttt{Lx\_final} & $\mathrm{erg\,s}^{-1}$ & X-ray luminosity used in the analysis, converted to ROSAT energy band  &8.32E+026 \\
 \texttt{Lx\_final\_uerr} & $\mathrm{erg\,s}^{-1}$ & upper error of \texttt{Lx\_final} & \\
 \texttt{Lx\_final\_lerr} & $\mathrm{erg\,s}^{-1}$ & lower error of \texttt{Lx\_final} & \\
 \texttt{Lx\_final\_upp\_lim} & & flag for \texttt{Lx\_final} as upper limit value & 0\\
 \texttt{L\_bol} & $\mathrm{erg\,s}^{-1}$ &bolometric luminosity & 5.83E+030\\
 \texttt{r\_L\_bol}&&reference of \texttt{L\_bol}& GAIA\tablefootmark{b}  \\
 \texttt{Lx/L\_bol} &  & $L_\mathrm{X}/L_\mathrm{bol}$ used in the final analysis & 1.43E$-04$
\\
 \texttt{Lx/L\_bol\_uerr} &  & upper error of \texttt{Lx/L\_bol} & \\
 \texttt{Lx/L\_bol\_lerr} &  & lower error of \texttt{Lx/L\_bol} & \\
 
\hline

\end{tabular}

\tablefoottext{a}{ROSAT all-sky survey also used PSPC, and we distinguish between \texttt{RASS} (survey) and \texttt{PSPC} (pointed observation).}

\tablefoottext{b}{The references are the \textit{Gaia} parameter catalog (\texttt{GAIA}), $L_\mathrm{bol}$ calculated with the \textit{Gaia} BC routine (\texttt{GAIA\_EEM}) and the Mamajek table (\texttt{EEM}).}

\end{table*}

\FloatBarrier 
\twocolumn

\end{appendix}

\end{document}